\documentclass[12pt,a4paper]{article}
\usepackage{epsf,epsfig,graphicx} 

\usepackage[normalem]{ulem}  

\usepackage{color}
\newcommand{\goo}{\,\raisebox{-.5ex}{$\stackrel{>}{\scriptstyle\sim}$}\,}
\newcommand{\loo}{\,\raisebox{-.5ex}{$\stackrel{<}{\scriptstyle\sim}$}\,}

\newcommand{\midtilde}{\raisebox{-0.25\baselineskip}{\textasciitilde}}

\title{A comparative study of statistical models for nuclear equation of state 
of stellar matter}
\begin{document}
\maketitle
\begin{center}
{\Large N.~Buyukcizmeci$^{a,b}$, A.S.~Botvina$^{b,c,d}$, I.N.~Mishustin$^{b,e}$, 
R. Ogul$^{a}$, 
M.~Hempel$^{f}$, J.~Schaffner-Bielich$^{g,}$\footnote{Present address: Institute for Theoretical Physics, J.W. Goethe-Universit\"at, Max-von-Laue-Stra$\ss$e 1, 60438 Frankfurt am Main, Germany}
, F.-K.~Thielemann$^f$,
S.~Furusawa$^{h}$, K.~Sumiyoshi$^{i,j}$, S. Yamada$^{h,k}$, H. Suzuki$^{l}$}
\end{center}
\begin{center}
{\it
 $^a$Department of Physics, Selcuk University, 42079 Kampus, Konya, Turkey\\
 $^b$Frankfurt Institute for Advanced Studies, J.W. Goethe University, D-60438
     Frankfurt am Main, Germany\\
 $^c$Institute for Nuclear Research, Russian Academy of Sciences, 
117312 Moscow,
 Russia\\
 $^d$Helmholtz Institute Mainz, J. Gutenberg University, 55099 Mainz, Germany\\
 $^e$Kurchatov Institute, Russian Research Center, 123182 Moscow, Russia\\
 $^f$Departement Physik, Universit\"at Basel, Klingelbergstr.~82, 4056 Basel, 
Switzerland\\ 
$^g$Institut f\"ur Theoretische Physik, Ruprecht-Karls-Universit\"at, 
Philosophenweg 16, 69120 Heidelberg, Germany\\
 $^h$Department of Science and Engineering, Waseda University, 3-4-1 Okubo, 
Shinjuku, Tokyo 169-8555, Japan\\
 $^i$Numazu College of Technology, Ooka 3600, Numazu, Shizuoka 410-8501, 
Japan\\
$^j$Theory Center,High Energy Accelerator Research Organization (KEK), 
Oho 1-1, 
Tsukuba, Ibaraki 305-0801, Japan\\
 $^k$Advanced Research Institute for Science and Engineering, Waseda 
University, 3-4-1 Okubo, Shinjuku, Tokyo 169-8555, Japan\\
 $^l$Faculty of Science and Technology, Tokyo University of Science, Yamazaki 
2641, Noda, Chiba 278-8510, Japan\\
}
\end{center}
\normalsize
\vspace{0.3cm}
\begin{abstract}
We compare three different statistical models for  the equation of state (EOS) 
of stellar matter at subnuclear densities and 
temperatures (0.5-10 MeV) expected to occur during the 
collapse of massive stars and supernova explosions. 
The models introduce the distributions of various 
nuclear species in nuclear statistical equilibrium, but use somewhat 
different nuclear physics inputs. It is demonstrated that the basic 
thermodynamical quantities of stellar matter under these conditions are 
similar, except in the region of high densities and low temperatures. 
We demonstrate that mass and isotopic distributions have considerable 
differences related to the different assumptions of the models on 
properties of nuclei at these stellar conditions. Overall, the 
three models give similar trends, but the details reflect the uncertainties 
related to the modeling of medium effects, such as the temperature and 
density dependence of surface and bulk energies of heavy 
nuclei, and the nuclear shell structure effects. We discuss importance of new physics inputs for astrophysical calculations from experimental data obtained in intermediate energy heavy-ion collisions, in particular, the similarities of the conditions reached during supernova explosions and multifragmentation reactions.
\end{abstract}
\vspace{0.3cm}
\vspace{0.3cm}

\section{Introduction}
\vspace{0.3cm}

It is known that the short-range strong interactions between 
nucleons can lead to a fast equilibration in violent nuclear reactions. 
Therefore, statistical models have proved 
to be very successful for interpretation of nuclear reactions at various 
energies. They are widely used for description of the fragment production 
when one or several equilibrated sources can be identified. 
Originally, this concept was proposed by Niels Bohr \cite{Bohr} for 
description of a compound nucleus decaying via evaporation of light 
particles or fission. 
Recently, it has been demonstrated that the concept of statistical 
equilibrium can be successfully used even for violent multifragmentation 
reactions leading to production of many intermediate mass 
fragments \cite{Randrup,Gross,SMM}. 

Moreover, the nuclear statistical equilibrium is established in many 
astrophysical processes, 
when the characteristic time for nuclear transformations is much shorter 
than those of these processes. For example, one of the 
most spectacular astrophysical events is a core-collapse supernova explosion, 
with a huge energy release of about 100~MeV per nucleon 
\cite{Brown,Bethe}. 
When the core of a massive star collapses, it may reach 
baryon densities which are several times larger than the normal nuclear 
density $\rho_0 \approx 0.15$ fm$^{-3}$ 
(i.e., $m_N \rho_0 \approx 2.5\cdot 10^{14}$~g/cm$^3$). 
The repulsive nucleon-nucleon interactions give rise to a
bounce 
of the central core and the creation of a shock wave propagating through
the 
in-falling stellar material. Many supernova simulations show that soon after the starting, the bounce shock turns into a stalled shock wave due to energy loss by dissociation of heavy nuclei in the in-falling material and by neutrino emission \cite{burrows2012, janka2012}. If, at a later evolution, this 
shock wave will revive and produce a matter flow with positive velocities, 
it will lead to the ejection of the star's envelope
observed as a supernova explosion. During the collapse and subsequent 
explosions, the temperatures and the densities reach $T\approx
(0.5\textrm{--}30)$~MeV and $\rho \approx (10^{-10}\textrm{--}3) \rho_0$, respectively. It is
expected that the nuclear 
statistical equilibrium is established under these conditions. 

As demonstrated by several studies (see, e.g., refs. 
\cite{Janka1,Janka,Thielemann,Sumi,Burrows}), 
present hydrodynamical simulations of core-collapse supernovae 
in spherical symmetry are not able to produce successful explosions 
for most of the progenitors, except for the smallest progenitor masses 
in the range of 8 to 10 M$_\odot$ \cite{kitaura06,fischer10}. 
Multidimensional effects like fluid instabilities, convection and rotation 
can successfully lead to explosions 
\cite{3d,takiwaki12,mueller12,winteler12,burrows07}, 
but presently the results of different groups have not yet converged. 
On the other hand, it is known that the nuclear composition, i.e., the 
mass fractions of nuclei and nucleons, is also extremely important 
for understanding the physics of supernovae. In particular, the 
weak reaction rates and energy spectra of emitted neutrinos are very sensitive 
to the presence of heavy nuclei (see, e.g., 
\cite{Ring,Hix,Langanke,Horowitz}). 
The nuclear distributions are especially relevant for weak reactions. 
For example, the new treatments of electron captures on heavy nuclei in 
Refs.~\cite{langanke03,langanke03b,hix03} and inelastic 
neutrino--nucleon (nuclei) scattering in Ref.~\cite{langanke08} are 
all based on the distributions of nuclei. The effect of nuclear composition can be even more important if the properties of nuclei (for example, their symmetry energy) embedded in supernova medium will be modified and this will lead to a dramatic change in the rates of neutrino reactions and the electron absorption \cite{Botvina10}. In this case the energy deposition from neutrino processes will essentially change, as well as the electron fraction, which will influence the dynamics of the collapse and explosion. 
Nuclear reactions in the supernova environment are also important, 
because supernova explosions may be considered as breeders for creating 
chemical elements, heavier than Fe and Ni. 
Pronounced peaks in the element abundances can be explained by neutron 
capture reactions in s- and r-processes \cite{Cowan,Qian}. 
It was always expected 
that suitable conditions for the r-process were provided by free neutrons 
abundantly produced in supernova environments together with appropriate 
seed nuclei. However, this is still a topic of current research, as 
typical core-collapse supernova simulations do not lead to a successful 
r-process. Alternatively, also neutron star mergers are considered as a 
source of r-process elements. 

One of the initial EOSs for
supernova matter, which is frequently 
used in supernova simulations, was proposed in Refs. \cite{Lamb,Lattimer} 
many years ago. It includes nucleons, alpha particles, 
and heavy nuclei in statistical equilibrium. 
It is obtained under the assumption that the whole ensemble of hot 
heavy nuclei, i.e., with mass number $A$ larger than
4, 
can be replaced by a single ``average'' nucleus. 
The same assumption was also used in the EOS within a relativistic 
mean-field (RMF) approach given in Ref. \cite{Shen}. 
Recent models \cite{Botvina10,FYSS,Japan,Botvina04,Hempel10,blinnikov11,gshen2010b}, 
include ensembles with various nuclear species, which are 
most important for the accurate description of the neutrino 
transport during supernova explosion processes, 
though the distribution of heavy nuclei may have only a minor direct 
effect on thermodynamic quantities at some particular conditions \cite{burrows84}. 

Presently, it is under discussion if only the nuclei in long-lived states known from 
terrestrial experiments should be included in this ensemble 
(see refs.~\cite{Japan,Langanke_structure,Janka2}), or 
particle-unstable states and new unknown exotic nuclei should be considered to have
a contribution, as well \cite{Botvina10,FYSS,Botvina04,Hempel10}. 
The last hypothesis is quite 
justified at temperatures higher than 1 MeV, when the nuclear shell 
structure is washed out and the liquid-drop description becomes more 
adequate\cite{Ignatiuk,K-H-Schmidt}. Moreover, at high temperature and large baryon densities $\rho_B \approx 10^{-3}\textrm{--}10^{-1} \rho_0$, the nuclei can still
interact with surrounding species and experience thermal expansion. The properties of nuclei in such environments may differ significantly from the ones obtained in low-energy nuclear experiments \cite{LeFevre,Souliotis,Ogul}. 
This is a challenging problem for both theoretical and
experimental nuclear physics.

Properties of strongly-interacting nuclear matter have been studied experimentally 
and theoretically for a long time. On a qualitative level, the phase diagram 
of symmetric and neutron-rich nuclear matter is understood rather well (see, 
e.g., refs. \cite{Lamb,Mosel76,Lamb78}). It is commonly accepted that 
the nuclear phase diagram contains a liquid-gas phase coexistence line with 
a critical temperature $T_c$ about 10--20~MeV. In many models this phase transition is of first-order, if only nuclear matter is considered, and Coulomb interactions are ignored, see e.g., \cite{Soma2009, Fiorilla2012}. Note that the formation of nuclei in astrophysical environment involves the Coulomb and other interactions existing in finite nuclei, where some features of the first order phase transition may be preserved. The critical temperature extracted from experiments on finite nuclei \cite{Karna03, Moretto11} is in the range $T_c \approx 17 - 20$ MeV. In the present work we consider explicitly the ensemble of nuclei embedded in the background of nucleons and electrons. Still the 
comparison with the nuclear matter phase diagram gives important
insight 
into the overall behavior of the supernova EOS. 

\begin{figure} 
\begin{center}
\includegraphics[width=10cm,height=10cm]{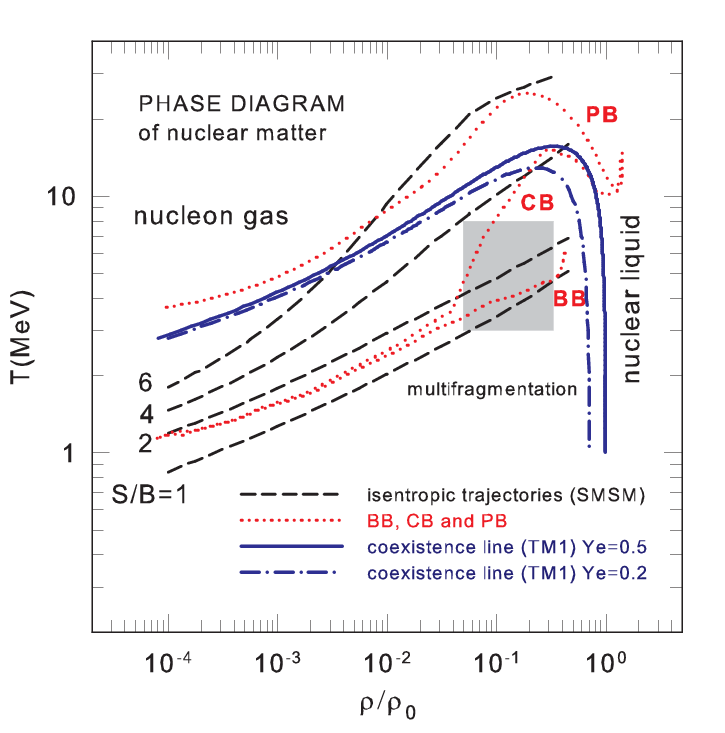} 
\end{center}
\caption{\label{fig_phase}\small{Nuclear phase diagram in the 'temperature -- baryon density' 
plane. Solid black and dashed-dotted purple lines indicate boundaries of 
the liquid-gas 
coexistence region for symmetric ($Y_e=0.5$) and asymmetric matter 
($Y_e=0.2$) calculated with TM1 interactions \cite{Suga94}. 
The shaded area corresponds to typical conditions for nuclear 
multifragmentation reactions \cite{Botvina10}. The dashed black lines 
are isentropic trajectories 
characterized by constant entropy per baryon, $S/B=$1, 2, 4 and 6 calculated 
with SMSM \cite{Botvina10}. The dotted red lines show model results of 
Ref.~\cite{Sumi} 
for BB (just before the bounce), CB
(at the core bounce) and PB (the post 
bounce) in a core-collapse supernova. (Color version online.) 
}}
\end{figure}

In order to compare the thermodynamical conditions obtained in nuclear 
reactions and in supernovae, we show the phase diagram for symmetric and 
asymmetric nuclear matter in Fig.~\ref{fig_phase}, for
a range of 
densities and temperatures appropriate for core-collapse supernova explosions. 
The electron fraction $Y_e$, which is equal to the total proton fraction, 
in the supernova core varies from 0.1 to 0.5. 
In the two-phase coexistence region (below the solid and dot-dashed
lines) at densities 
$\rho \approx 0.3-0.8\rho_0$ the matter should be in a 
mixed phase, which  is strongly  inhomogeneous with
intermittent dense and dilute regions. 
In the coexistence region at lower densities, $\rho <
0.3\rho_0$ 
the nuclear matter breaks up into compact nuclear droplets surrounded by 
nucleons. These relatively low 
densities dominate during the main stages of stellar collapse and explosion. 
Under such conditions one can expect a mixture of nucleons and light and 
heavy nuclei. 

In Fig.~\ref{fig_phase} we demonstrate also isentropic trajectories in 
nuclear matter (dashed curves) and the trajectories of density and 
temperature inside the supernova core, which are taken from the supernova 
simulation of the star with 15 solar masses 
\cite{Sumi} (red dotted curves). 
The snapshots in the supernova dynamics are selected 
for three stages before, at, and after the core bounce. 
In the gravitational 
collapse from the iron core of the star (BB: just before
bounce), 
the density and temperature roughly follows the isentropic 
curves S/B=1--2. At the core bounce, when the central density 
increases just above the nuclear matter density $\rho_0$, 
the temperature of the inner core becomes higher than 10 MeV 
due to the passage of the shock wave (CB: core bounce). 
The temperature of the whole supernova core is still high 
at 150 ms after the core bounce (PB: post bounce).  
The shock wave is stalled around 130 km in this 1D 
calculation, which does not lead to an explosion. 

These conditions inside the supernova core pass through interesting regions of the phase diagram. 
The BB and CB trajectories go through the multifragmentation region, 
which motivates us to assess the supernova conditions 
by the multifragmentation reactions taking place in heavy-ion
collisions 
of intermediate energy. 
The trajectories after the bounce (CB and PB) traverse 
the phase boundary between the mixture of nuclei and 
the nucleon gas. It is also interesting that 
this region is dominated by light nuclei ($^4$He and lighter ones) 
\cite{Hempel10,Hempel11,typel10,Hempel11b,Sumi08,Arcones08}.
Since the dynamics of the shock wave is affected by 
the interaction of neutrinos with nucleons and nuclei, 
it is important to determine the composition of hot dense 
matter in this region.  In fact, the heating through 
neutrino absorption contributes to 
the revival of shock wave together with the hydrodynamical 
instabilities in 2D and 3D supernova simulations.  
Large abundance of light nuclei ($A=2, 3$) may contribute to the 
modifications of heating rate and/or energy spectrum through neutrino 
reactions with these nuclei \cite{OConnor07,nakamura09}. 

Our goal is to compare three different models for the EOS of stellar
matter 
at subsaturation densities. The calculations are done for wide intervals 
of $T$, $\rho$ and $Y_e$. 
All the three EOS models investigated in the present 
article, contain 
detailed information about the nuclear composition as they are built 
from a statistical distribution of different nuclear species. 
In the present paper we do not try 
to give preference to any of them, by keeping in mind that properties
of 
nuclei in stellar medium and interactions between nuclear species in 
diluted matter are not known sufficiently well. 
We rather want to identify how the different model assumptions affect 
thermodynamic quantities and the nuclear composition. In the future we plan 
to construct a unified EOS which includes the 
components of these
models 
which are best verified by theoretical and experimental studies. 
We believe that this kind of comparison of 
the EOSs will be very useful for people working on
nuclear astrophysics. 
We clearly demonstrate the similarities and 
differences of the model results, and discuss their physical reasons. It is 
necessary to mention that all models treat electrons and photons in the 
same way, and differences come entirely due to a different description 
of nuclei and the underlying nuclear
interactions. Therefore, if the photon and lepton contributions 
dominate, all models give similar results.

This article is organized as follows. In section 2, we explain the formalism 
of the three EOSs respectively. 
The comparison of results is presented in section 3, where we show 
predicted mass and isotope distributions, their moments and fractions, as 
well as thermodynamic quantities. 
The paper is wrapped up with a summary and some discussions in section~4. 

\vspace{0.3cm}
\section{Statistical models for supernova matter}
\vspace{0.3cm}

First, we should remark about experimental possibilities for studying fragmented nuclear matter at subsaturation density. Traditionally, only theoretical approaches which use nuclear forces extracted from experimental study of nuclear structure were applied for this purpose. However, as became clear after last 20 years of intensive experimental studies, many nuclear reactions lead to the formation of thermalized nuclear systems characterized by subnuclear densities and temperatures of 3-8 MeV. De-excitation of such systems goes through nuclear multifragmentation, i.e. break-up into many excited fragments and nucleons. We emphasize that thermal and chemical equilibrium can be established in many multifragmentation reactions, and it is generally accepted by the scientists involved in the field. Transport theoretical calculations of central heavy-ion collisions around the Fermi-energy (i.e., with energies of 20--50 MeV per nucleon) predicts momentum distributions of nucleons which are similar to equilibrium ones after first $\approx 100$ fm/c (see, e.g., \cite{Botvina1995} and other references in the topical issue \cite{EPJA}). In peripheral heavy-ion collisions there is a midrapidity region which can be associated with a dynamical fragment formation, however, the projectile and target excited residues represent thermal sources, which later on expand and break-up into many fragments. The final proof of equilibration was provided by numerous comparisons of statistical models with experimental data: These models perfectly describe many characteristics of nuclear fragments observed in multifragmentation experiments: multiplicities of intermediate mass fragments, charge and isotope distributions, event by event correlations of fragments (including fragments of different sizes), their angular and velocity correlations, and other observables \cite{Gross,SMM,Souliotis,Ogul,Botvina90,ALADIN,EOS,MSU,INDRA,FASA,Dag,Hauger,Iglio, Hudan,Wang,Viola,Rodionov,Pienkowski}. The parameters of nuclear matter at the moment of formation of hot fragments, such as temperature and density, can also be established reliably in experiment by using observed relative velocities of fragments, and relative isotope yields \cite{EPJA}. Typical conditions associated with multifragmentation reactions are indicated by the shaded area in Fig.~\ref{fig_phase}. These reactions give us a chance to access hot nuclei in the environment of other nuclear species in thermodynamical equilibrium as we expect in supernova matter. The properties of these nuclei can be directly extracted from experimental data and then this information can be used for more realistic calculations of nuclear composition in stellar matter. As one can estimate from Fig.~\ref{fig_phase}, in the course of massive star collapse the stellar nuclear matter passes exactly through the multifragmentation region with typical entropy per baryon $S/B=1-4$.

Below we consider three typical models for supernova matter, which are based 
on the 
assumption of statistical equilibrium: The Statistical Model for Supernova 
Matter (SMSM) 
\cite{Botvina10,Botvina04}, the statistical model of 
Hempel and Schaffner-Bielich (HS) \cite{Hempel10} and 
the statistical model of Furusawa, Yamada, Sumiyoshi and 
Suzuki (FYSS) \cite{FYSS}. In this section we briefly 
describe the models emphasizing their similarities and differences. 

\vspace{0.3cm}
\subsection{EOS -- SMSM}
\vspace{0.3cm}

This model is a generalization of well-known statistical multifragmentation model (SMM) \cite{SMM} for astrophysical conditions. The SMM is one of the most successful models used for the theoretical description of multifragmentation reactions as demonstrated in the above cited publications,and it can be easily extended to describe clusterized stellar matter, what is not possible with other statistical models, e.g., MMMC \cite{Gross}, designed for finite nuclear systems only.

We give a brief description of the SMSM's nuclear part, which is important 
for comparison with other models. The full description of the model including 
electron, photon, and neutrino contributions (in thermal equilibrium) 
one can find in Ref.~\cite{Botvina10}. Generally, the system is characterized 
by the temperature $T$ and baryon density $\rho_B$ and electron fraction 
$Y_e=\rho_e/\rho_B$, where $\rho_e$ is the net electron density. 
The nuclear component of supernova matter is represented as a mixture of 
gases of different species $(A,Z)$ including nuclei and nucleons. It is 
convenient to introduce the numbers of particles of different kind $N_{AZ}$ 
in a normalization volume $V$. Then the total free energy density of this 
mixture can be written as 
\begin{eqnarray} \label{energy}
f&=&\frac{1}{V}\sum_{AZ} N_{AZ} \left\{- T\cdot \left[ 
\ln \left(\frac{g_{AZ}^0 V_f A^{3/2}}{N_{AZ} \lambda_T^3}\right)+1\right]+
F_{AZ}\right\} 
\end{eqnarray}
The first term comes from the translational motion of particles, and the 
second term is associated with the binding energy and excitation of nuclear 
fragments ($A>4$). $g_{AZ}^0$ is the ground-state degeneracy factor of 
species $(A,Z)$, $\lambda_T=\left(2\pi\hbar^2/m_NT\right)^{1/2}$ is the 
nucleon thermal wavelength, $m_N \approx 939$ MeV is the average nucleon 
mass. $V_f$ is the so-called free volume, which accounts for the finite
size of 
nuclear species. The SMSM does not include excited states explicitly, but 
incorporates a temperature dependence of the nuclear free energy $F_{AZ}$ 
of nuclei as will be presented below. 
We assume that all nuclei have normal nuclear density
$\rho_0$, so that the proper volume of a nucleus with
mass $A$ is $A/\rho_0$. At low densities the finite-size corrections can be 
included via the excluded volume approximation, 
$V_f/V \approx \left(1-\rho_B/\rho_0\right)$.
This approximation is commonly accepted in statistical models, 
and it is considered as a reasonable one at densities 
$\rho_B < 0.1 \rho_0$. 
Some information about the free volume at higher densities can be extracted 
from analysis of experimental data obtained in multifragmentation reactions 
\cite{EOS}. Eq.~(\ref{energy}) can also be written in the following form:
\begin{equation}
f= \frac{1}{V}\sum_{AZ} N_{AZ} \left\{F^{t0}_{AZ}-T\ln(V_f/V)+F_{AZ} \right\}, 
\label{fsmsm}
\end{equation}
where
\begin{equation}
F^{t0}_{AZ} = - T\cdot \left[ 
\ln \left(\frac{g_{AZ}^0 V A^{3/2}}{N_{AZ} \lambda_T^3}\right)+1\right]
\end{equation}
corresponds to the translational energy of an ideal gas without the excluded 
volume correction. 

The numbers of nuclear species are constrained by the conditions
for baryon number
conservation and electro-neutrality
\begin{eqnarray}
\label{rhotot}
\rho_B=\frac{1}{V}\sum_{AZ}AN_{AZ}~,~
\rho_Q=\frac{1}{V}\sum_{AZ}ZN_{AZ}-\rho_e=0~.
\end{eqnarray}
We have performed calculations within the Grand Canonical approximation.
In this case the conservation laws (\ref{rhotot}) are fulfilled only for the 
mean values
$<N_{AZ}>$. These values are obtained by minimizing the free energy 
(\ref{energy}) under
constraints (\ref{rhotot}). The resulting expression is 
\begin{eqnarray} \label{naz}
<N_{AZ}>=g_{AZ}^0V_f\frac{A^{3/2}}{\lambda_T^3}
{\rm exp}\left[-\frac{1}{T}\left(F_{AZ}-\mu_{AZ}\right)\right]~,
\end{eqnarray}
where
\begin{equation} \label{chem}
\mu_{AZ}=A\mu_B+Z\mu_Q
\end{equation}
is the chemical potential of species $(A,Z)$, $\mu_B$ and $\mu_Q$ are chemical 
potentials responsible for 
the baryon number and charge conservation. For protons $\mu_p=\mu_B+\mu_Q$, 
for neutrons $\mu_n=\mu_B$. Below we use average densities 
$\rho_{AZ}=<N_{AZ}>/V$.

The internal excitations of nuclei play an important role in regulating 
fragment's abundance, since they increase significantly their entropy. 
We calculate the internal 
excitation energy of nuclei by assuming that they have the same internal 
temperature as the surrounding medium. In this case not only 
particle-stable states but also particle-unstable states will contribute 
to the excitation energy and entropy. This assumption can be justified by the 
dynamical equilibrium of nuclei in the hot environment \cite{Botvina10}, 
and is supported by both many comparisons of SMM with experimental data on 
multifragmentation and direct experimental measurements \cite{hudan03}. 
Moreover, in the supernova environment both the excited states and the 
binding energies of nuclei will be strongly affected by the surrounding 
matter. By this reason, we find it more appropriate to use an approach 
which can easily be generalized to include in-medium modifications of 
nuclear properties. 
Namely, the internal free energy of species $(A,Z)$ with $A>4$ is
parameterized in the spirit of the liquid-drop model, which has been 
proved to be very successful in nuclear physics \cite{Bohr,SMM}:
\begin{equation}
F_{AZ}(T,\rho)=F_{AZ}^B+F_{AZ}^S+F_{AZ}^{\rm sym}+F_{AZ}^C~~. \label{faz_smsm}
\end{equation}
Here the right-hand side contains, respectively, the bulk,
the surface, the symmetry and the Coulomb terms. The first two terms
are temperature-dependent and are motivated by properties of nuclear 
matter corresponding to the liquid-gas phase transition \cite{SMM}, the third 
one is temperature independent: 
\begin{eqnarray}
F_{AZ}^B(T)=\left(-w_0-\frac{T^2}{\varepsilon_0}\right)A~~, 
\label{fbul}\\
F_{AZ}^S(T)=\beta_0\left(\frac{T_c^2-T^2}{T_c^2+T^2}\right)^{5/4}A^{2/3}~~, 
\label{fsuf}\\
F_{AZ}^{\rm sym}=\gamma \frac{(A-2Z)^2}{A}~~.
\label{fsym}
\end{eqnarray}
Here $w_0=16$ MeV, $\varepsilon_0=16$ MeV, $\beta_0=18$ MeV, 
$\gamma=25$ MeV, and $T_c=18$ MeV are the model parameters which are 
extracted from nuclear phenomenology and provide a good description of 
multifragmentation data \cite{SMM,ALADIN,EOS,MSU,INDRA,FASA,Dag}. 
However, these parameters can be easily adopted to the new conditions 
expected in stellar environment. Nucleons and light fragments with A$\leq$4 are considered as structure-less 
particles characterized only by exact masses and proper volumes \cite{SMM}. 
For them we adopt $F_{AZ}=-B_{AZ}+F_{AZ}^C$, where $B_{AZ}$ is 
the measured binding energy. 
In the electrically-neutral environment the fragment Coulomb energy 
$F_{AZ}^C$ is modified by the screening effect of electrons. 
In the SMSM it is calculated by using the Wigner-Seitz approximation 
\cite{Botvina10,Lamb} for all fragments and charged particles (including 
protons). 
\begin{eqnarray}
F_{AZ}^C(\rho)=\frac{3}{5}c(\rho)\frac{(eZ)^2}{r_0A^{1/3}}~~,\\
c(\rho)=\left[1-\frac{3}{2}\left(\frac{\rho_e}{\rho_{0p}}\right)^{1/3}
+\frac{1}{2}\left(\frac{\rho_e}{\rho_{0p}}\right)\right]~, \label{fazc}
\end{eqnarray}
where $r_0=1.17$ fm and $\rho_{0p}=(Z/A)\rho_0$ is the proton
density inside the nuclei. The screening function $c(\rho)$ is 1
at $\rho_e=0$ and 0 at $\rho_e=\rho_{0p}$. 
To simplify calculations one can use an approximation 
$\rho_e/\rho_{0p}=\rho_B/\rho_0$, as in ref.~\cite{Lamb}, which works well 
when neutrons are mostly bound in nuclei, and leads to very similar results 
in many cases. The average densities of all nuclear species are calculated 
self-consistently by taking into account the relations between their 
chemical potentials. We perform calculations for all fragments with 
1$\leq A \leq$1000 and 0$\leq Z \leq A$. This restriction on the size of 
nuclear fragments is fully justified in our case, since fragments with 
larger masses ($A>1000$) can be produced only at very high densities 
$\rho \goo 0.3\rho_0$ \cite{Bethe,Lamb}, which are appropriate for 
regions deep inside of protoneutron stars, and which are not
considered 
in this model. The SMSM EOS tables (like Shen \cite{Shen} or HS \cite{Hempel10}) are currently under construction for publication, and will 
be available in the Internet soon.\footnote{See \texttt{http://fias.uni-frankfurt.de/physics/mishus/research/smsm/}} 

\vspace{0.3cm}
\subsection{EOS --  HS}
\vspace{0.3cm}

In the HS model for the EOS, matter is described as an ensemble of
nucleons 
and nuclei in nuclear statistical equilibrium (NSE), whereas interactions 
of the nucleons and excluded volume corrections are implemented. We 
remark that tabulated versions of the EOS-HS have recently been applied 
in core-collapse supernova simulations \cite{Hempel11}. These tables 
are available online for five different parameterizations of relativistic 
mean-field (RMF) interactions.\footnote{See \texttt{http://phys-merger.physik.unibas.ch/\midtilde hempel/eos.html}\label{eospage}.} 
In Ref. \cite{Hempel11b} it was shown that the HS model gives a similar 
description of the medium effects on light clusters like two quantum 
many-body models. 
This model is in good agreement with experimental data for the equilibrium 
constant $K_c$ at densities between 0.02 and 0.03 fm$^{-3}$ and 
temperatures around 10 MeV \cite{Qin}. 
In the 
following, we give a brief summary of the HS model, all details can be 
found in Ref. \cite{Hempel10}.

In the HS model nuclei are treated as non-relativistic classical particles 
with Maxwell-Boltzmann statistics. For the description of nuclei the 
experimentally measured masses from the atomic mass table 2003 from Audi, 
Wapstra, and Thibault \cite{Audi} are used. Shell effects are thus 
naturally included. In addition, for the masses of experimentally unknown 
nuclei we use results of theoretical nuclear structure calculations 
via the nuclear mass table of Geng et al. \cite{Geng05}. We have chosen this mass table because it was calculated with the TMA interactions, which are very similar to the TM1 interaction used in the paper for the unbound nucleons. Other mass tables can also be used, however, we have found that the selection of the mass table which provides very similar ground state masses (with the differences around $\approx 1$ MeV) does not change the overall behavior of the EOS model and its main characteristic features. This mass table 
lists 6969 even-even, even-odd and odd-odd nuclei, extending from 
$^{16}_{~8}$O to $^{331}_{100 }$Fm from slightly above the proton to 
slightly below the neutron drip line. However, nuclei beyond the neutron 
drip line have been excluded in the EOS calculation of HS.
This is done 
because nuclear structure calculation are not very reliable beyond the 
neutron drip-line, and to have a clear physical criteria which nuclei to 
be included. The total binding energy of nucleus 
$(A,Z)$ will be denoted by $B_{AZ}$ hereafter. The total mass of a nucleus 
is set by its rest-mass and 
its binding energy, $M_{AZ} = M_{AZ}^0 - B_{AZ}$. 

Differently than in SMSM, temperature effects in HS are implemented 
via a temperature-dependent internal partition function $g_{AZ}(T)$. It 
represents the sum over all excited states of a hot nucleus. The 
semi-empirical expression from Ref.~\cite{Fai82} is used: 
\begin{eqnarray}
&g_{AZ}(T)=g_{AZ}^{0} +\frac{0.2}{A^{5/3}\textrm{MeV}}\int_0^{E_{\rm max}} dE^*
e^{-E^*/T}\exp\left(\sqrt{2 a(A) E^*}\right)&, \label{eq_gaz} \\
& a(A)=\frac A 8 (1-0.8A^{-1/3})\rm {~MeV} ^{-1} & \nonumber
\end{eqnarray}
with $g_{AZ}^0$ denoting the spin-degeneracy of the ground-state, as before 
in SMSM. In the
original reference of the HS model \cite{Hempel10} 
$E_{AZ}^{max}= \infty$ was chosen. For the supernova simulations in 
Ref. \cite{Hempel11} HS changed the value of $E_{AZ}^{max}$ to the 
binding energy of the nucleus, $E_{AZ}^{max}=B_{AZ}$, which is also 
used here. It means that only excited states are considered which are 
still bound, to avoid an overestimation of internal excitation energies 
at large temperatures.
It is well known that in the low temperature limit Eq.(\ref{eq_gaz})
leads to 
the $T^2$ term in the free energy, as also assumed in the SMSM, 
Eq.(\ref{fbul}). However, in the SMSM the level density parameter $a$ is 
taken about twice smaller than the empirical value ($\approx
A/8$~MeV$^{-1}$) 
because of the additional contribution from the surface term,
Eq.(\ref{fsuf}). 

In the HS model, nuclear matter is described as a chemical mixture of
different 
nuclear species and nucleons. 
To assure the disappearance of nuclei above saturation density $\rho_0$, 
an excluded volume approach is used. Like in SMSM, this is done via the 
free volume fraction $V_f/V$. However, in HS the treatment of unbound 
nucleons is different from Boltzmann particles as taken in SMSM. In HS and also in FYSS, interactions 
of the unbound nucleons are taken into account with an RMF model. The 
RMF parameter set TM1 was chosen \cite{Suga94} in order to use the same 
approach for the nucleons as in the EOS of Shen et al. \cite{Shen}. 
Nucleons are assumed to be situated outside of nuclei, and therefore the 
filling factor of the nucleons $\eta=1-\sum_{AZ} A\rho_{AZ}/\rho_{0}$ is 
introduced. It relates the total number densities of neutrons $\rho_n$ and 
protons $\rho_p$, respectively, with the local number densities outside of 
nuclei $\rho'_n$ and $\rho'_p$, by $\rho_{n/p}=\rho'_{n/p} \eta$. 

Based on these assumptions, HS derive the following free energy density $f$: 
\begin{eqnarray}
f&=&\eta f^{RMF}(T,\rho_n',\rho_p') + \sum_{A>1,Z} \rho_{AZ}\left\{F_{AZ}^{t0}
-T \ln(V_f/V) + F_{AZ}\right\} \; . \label{f_hs}
\end{eqnarray}
The part with the sum over experimentally known and theoretically calculated 
nuclei $(A,Z)$ with $A>1$, has formally the same structure as in the free 
energy density of the SMSM (Eq.~(\ref{fsmsm})). Still  there are some 
important differences, which we want to discuss now. $F_{AZ}^{t0}$ is the 
translational free energy of a heavy nucleus: 
\begin{eqnarray}
F_{AZ}^{t0}=- T \cdot \left[ \ln \left(\frac{g_{AZ}(T) }
{\rho_{AZ}\lambda_{AZ}^3}\right)+ 1 \right], \; 
 \lambda_{AZ} = \left(2\pi \hbar ^2/M_{AZ} T  \right)^{1/2} \; ,
\end{eqnarray}
which is similar as in the SMSM, with the negligible difference that the 
masses $M_{AZ}$ appear instead of $A m_N$. However, the use of $g_{AZ}(T)$ 
in HS instead of $g^0_{AZ}$ in SMSM is an important difference, because 
$g_{AZ}(T)$ carries the main temperature dependence of heavy nuclei in HS. 
The excluded volume term $-T\ln(V_f/V)$ in Eq.~(\ref{f_hs}) is the same as 
in the SMSM. The internal free energy of the nucleus $(A,Z)$,
\begin{equation}
 F_{AZ}=-B_{AZ}+F_{AZ}^C \; ,
\end{equation}
is significantly different compared to the liquid-drop formulation of the 
SMSM (see Eq.~(\ref{faz_smsm})), mainly because of the binding energies 
$B_{AZ}$ which are based on experimental data and nuclear structure 
calculations. In SMSM the free energy includes a temperature
dependence 
in agreement with properties of matter in the region of nuclear liquid-gas 
phase transition. 
The Coulomb energy $F_{AZ}^C$ is described as in the SMSM, 
except that the constant part of the heavy nucleus is already included 
in $B_{AZ}$. The contribution of unbound nucleons in HS (the first term 
in Eq.~(\ref{f_hs})) is described separately by the RMF model with excluded 
volume corrections. Here the filling factor $ \eta$ appears in front of 
the pure RMF contribution of the nucleons $f^{RMF}$ which is set by their 
local number densities. The filling factor $\eta$ and the other excluded 
volume term $-T\ln(V_f/V)$ play an important role in the HS model, because 
they assure the disappearance of nuclei and a continuous transition to 
uniform nucleon matter at high densities.
The abundances of all nuclei and unbound 
nucleons are determined by the chemical equilibrium condition (\ref{chem}) 
and the conservation laws (\ref{rhotot}).  All other thermodynamic variables 
are then derived from Eq.~(\ref{f_hs}) in a thermodynamic consistent way. 

\vspace{0.3cm}
\subsection{EOS -- FYSS} \label{FYSSdis}
\vspace{0.3cm}

The formulation of the FYSS model is based on the NSE description using the 
mass formula for nuclei up to the atomic number of 1000 under the influence of 
surrounding nucleons and electrons. The mass formula is based on experimental 
data on nuclear binding energies that allow us to take into account nuclear 
shell effects. An extended liquid-drop model is used to describe the medium 
effects, and in particular, formation of the pasta phases. Because of this 
combination, the free energy of a multi-component system can reproduce the 
ordinary NSE results at low densities and make a continuous transition to 
the EOS for supra-nuclear densities. The details are given in Ref. 
\cite{FYSS}. FYSS model is being improved in the important points 
which appear in this comparison work.

Below, we give a short description of FYSS EOS. Assuming NSE, the abundances 
of nuclei as a function of $\rho_B$, $T$ and $Y_p$ are calculated by 
minimizing the model free energy with respect to the number densities of 
nuclei and nucleons under the constraints (\ref{rhotot}), and
(\ref{chem}). The free energy density of FYSS consists of contributions from 
unbound nucleons outside nuclei and the summation of translational 
$F_{AZ}^t$, bulk $F_{AZ}^B$, Coulomb $F_{AZ}^C$ and surface energies 
$F_{AZ}^S$ of all nuclei $(Z \leq 1000, N \leq 1000)$. 

The free energy density of the unbound nucleons, $f_{p,n}$, is calculated 
by the RMF with TM1 parameter set, which is the same as Ref. \cite{Shen}. 
FYSS takes into account the excluded-volume effect of free nucleons through 
$f_{p,n}=\eta f^{RMF}(T,\rho_n',\rho_p')$. This formula of the free energy 
density of the unbound nucleons is the same as in the HS model. 
On the other hand, the excluded volume effect for nuclei is different from 
that of the HS and SMSM models. FYSS assumes the nuclear translational motion 
contribution is calculated from Maxwell-Boltzmann statistics, however, the 
translational free energy of nuclei are suppressed by an additional volume 
factor $F_{AZ}^{t}=F_{AZ}^{t0} V_f/V$. Furthermore, $F_{AZ}^{t0}$ is slightly 
different from the expression in HS, because the mass term appearing in the 
thermal wavelength $\lambda_{AZ}$ contains liquid-drop modifications 
\cite{FYSS}. 
The contribution from the excited states of nuclei to the free energy is 
included in the $F_{AZ}^{t0}$ through $g_{AZ}(T)$. 
This is the same formula as the HS model but the upper limit of the integral 
is set to infinity, $E_{AZ}^{\rm max}=\infty$, being different from the 
HS model. 

To obtain the bulk energy of the nuclei, the experimental mass data 
\cite{Audi} are used at low densities whenever available. 
These experimental mass data are the same as in HS, but the theoretical 
data \cite{Geng05} included in HS are not used in FYSS. 
At high densities, where the nuclear structure is affected by the presence 
of other nuclei, nucleons and electrons, 
the bulk energy of the nuclei is approximated  by interpolation of the 
value obtained experimentally and the value derived theoretically from 
the RMF  between $10^{12} \rm{g/cm^3}$ and the nuclear saturation 
density, $\sim 10^{14.2} \rm{g/cm^3}$. 
The experimental and theoretical bulk energies are combined by the relation 
$F_{AZ}^{B} =M_{AZ}-[F_{AZ}^{C}]_{vacuum}-[F_{AZ}^{S}]_{vacuum}$ 
and $F_{AZ}^{B} =A F^{RMF}(\rho_{0AZ},T,Z/A)$ 
where $\rho_{0AZ}$ is the saturation density of the nucleus.
$F^{RMF}(\rho_B,T,Y_p)  $ is the free energy per baryon predicted by the RMF. 
 $\rho_{0AZ}(T)$ is set to the saturation density to have the lowest free 
energy of the RMF theory  $F^{RMF}(\rho_B,T,Z/A)$ for given $T$ and $Z/A$. 
More neutron-rich nuclei  at higher temperature have lower saturation density. 
Note that the proton fraction in this expression is not the one for the whole 
system 
but the one for each nucleus and that this bulk energy includes the symmetry 
energy.  
For very heavy and/or very neutron-rich nuclei with no experimental mass 
data available, the RMF is used for the evaluation of the bulk energy  at 
any density. 

As in SMSM and HS models, the Coulomb energy of nuclei is calculated using 
the Wigner-Seitz approximation. 
The outside unbound protons are 
included 
in the charge neutrality condition in addition to bound protons inside 
nuclei and uniformly distributed electrons. 

The surface energy of nuclei is given by the product of the nuclear 
surface area and the surface tension. 
\begin{eqnarray}
F_{AZ}^S=4\pi  R_{AZ}^{2}  \, 
\sigma_{AZ}  \left(1-\displaystyle{\frac{\rho'_p+\rho'_n}{\rho_{0}}} 
\right)^2 \\   
\sigma_{AZ}=\sigma_0  - \frac {A^{2/3} } {4 \pi R_{AZ}^2} 
[S_s(1- 2Z/A)^2 ],
\end{eqnarray} 
where  $R_{AZ} = ( 3/4 \pi V_{AZ})^{1/3} $ is the radius of nucleus 
$(A,Z)$ at baryon density $\rho$ and $ \sigma_0$ denotes the surface 
tension for symmetric nuclei. 
The surface tension  $\sigma_{AZ}$  includes the surface symmetry energy, 
where  neutron-rich nuclei have lower surface tensions than the symmetric 
nuclei. The values of the constants, $\sigma_0=1.15 \rm{MeV/fm^3}$ and 
$S_s =45.8 \rm{MeV}$, are adopted from Ref. \cite{Lattimer}. 
The last factor, $ \left(1-(\rho'_p+\rho'_n)/\rho_{0} \right)^2 $, is 
introduced to take into account the effect that the surface energy should be 
reduced as the density contrast decreases between the nucleus and the nucleon 
vapor. This surface energy has a dependence on the neutron-richness of nuclei 
and density of outside unbound nucleons. Note that contrary to the SMSM 
this model has no temperature dependence of the surface free energy. 

The free energy density of the FYSS model is 
\begin{equation}
 f = \eta f^{RMF}(T,\rho_n',\rho_p') + \sum_{AZ} \rho_{AZ} \{ F_{AZ}^{t0} 
V_f/V +F_{AZ} \} \; . \label{f_fyss}
\end{equation} 
The last term in the summation is the internal free energy of
the 
nucleus $(A,Z)$; 
\begin{equation}
 F_{AZ}=F_{AZ}^{B} +F_{AZ}^{C} +F_{AZ}^{S}.
\end{equation}
The bulk energies include the symmetry energies as previously stated and 
this internal free energies $F_{AZ}$ are equal to the experimental mass of 
the nucleus in the vacuum limit. 
Other thermodynamical quantities can be calculated from partial 
derivative of the minimized free energy explained above. 
In brief, the special feature of this model is to include in the calculations 
nuclear shell effects and very heavy nuclei $(Z > 100)$ 
by using the experimental mass data and the theoretical mass formula. 
Furthermore, FYSS assumes that each nucleus enters the nuclear pasta phase 
individually when the volume fraction in the Wigner-Seitz cell 
$u_{AZ}\simeq (A/\rho_{0AZ}(T))/(Z/\rho_e)$, i.e., the ratio of 
the nuclear volume by the cell volume, reaches $0.3$ and that the bubble 
phase is realized when it exceeds $0.7$. The intermediate states 
($ 0.3<u_{AZ}<0.7 $) are simply interpolated as other pasta phases. 
Under this assumption more neutron-rich nuclei go
into the pasta phases at 
lower densities than symmetric nuclei, since the volume fractions of
neutron-rich nuclei $u_{AZ}$ are lower than that of symmetric nuclei.
At high temperatures, the saturation densities  $\rho_{0AZ}(T)$ are low and 
the volume fractions $u_{AZ}$  are large compared with $u_{AZ}$  at low 
temperature and the same density.

\vspace{0.3cm}
\section{Comparison of results}
\vspace{0.3cm}

In this section we compare predictions of the three models for 
the nuclear composition and general thermodynamical properties of 
stellar matter. We present the results of
SMSM, HS 
and FYSS for the EOS at baryon densities
$\rho/\rho_0=10^{-3},10^{-2}$ and 
$10^{-1}$, and temperatures $T= $0.5--10 MeV. Since the full 
$\beta$-equilibrium is unlikely to be established in a supernova, we have adopted the
fixed electron fractions $Y_e$=0.2 and 0.4, 
which are typical for this scenario. The calculations
with $\beta$-equilibrium, which may be appropriate for a neutron star crust, 
one can find, e.g., in Refs. \cite{Botvina10,Botvina04}. 

\vspace{0.3cm}
\subsection{Mass distributions}
\vspace{0.3cm}

\begin{figure}  
\begin{center}
\includegraphics[width=12cm,height=16cm]{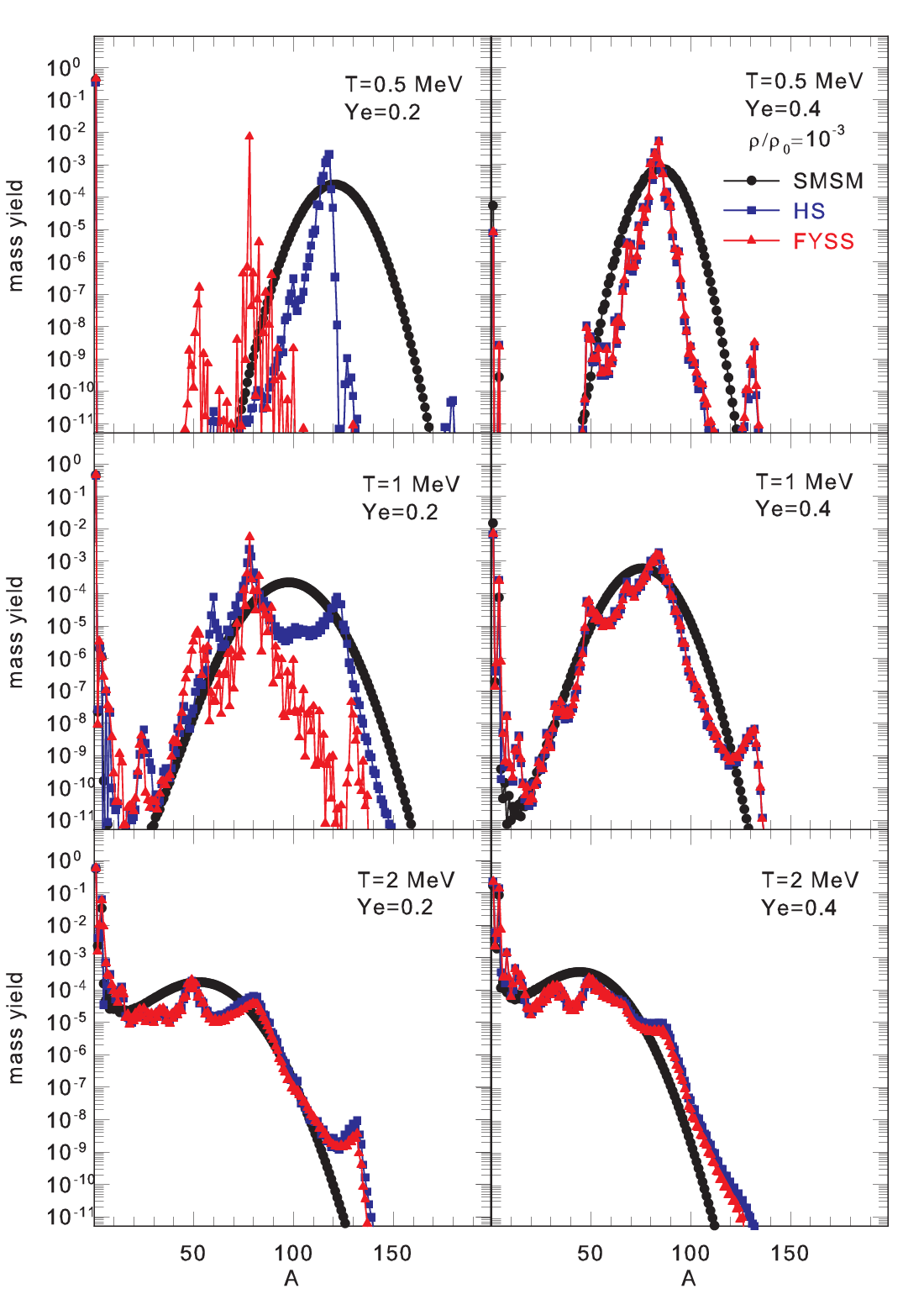}
\end{center}
\caption{\label{fig_ya1}\small{Mass distributions of fragments produced in matter with 
temperatures $T=0.5, 1$ and 2 MeV, electron
fractions $Y_e=0.2$ and 0.4, and  density $\rho/\rho_0=10^{-3}$. (Color version online.)
 }}
\end{figure}

\begin{figure}  
\begin{center}
\includegraphics[width=12cm,height=16cm]{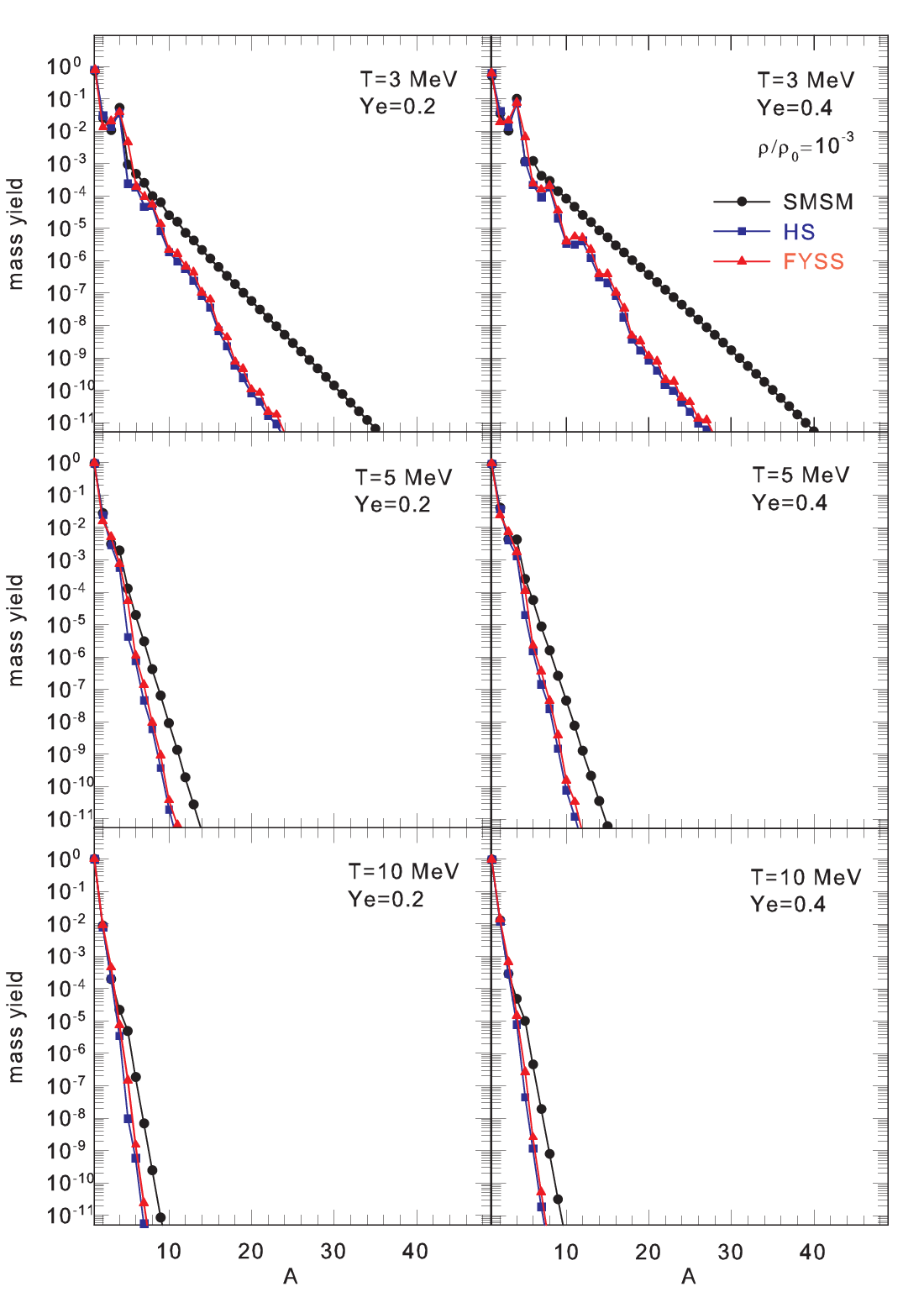}
\end{center}
\caption{\label{fig_ya2}\small{Mass distributions of fragments produced in matter with 
temperatures $T=3, 5$ and 10 MeV, electron
fractions $Y_e=0.2$ and 0.4, and  density $\rho/\rho_0=10^{-3}$. (Color version online.)
 }}
\end{figure}

\begin{figure}  
\begin{center}
\includegraphics[width=12cm,height=16cm]{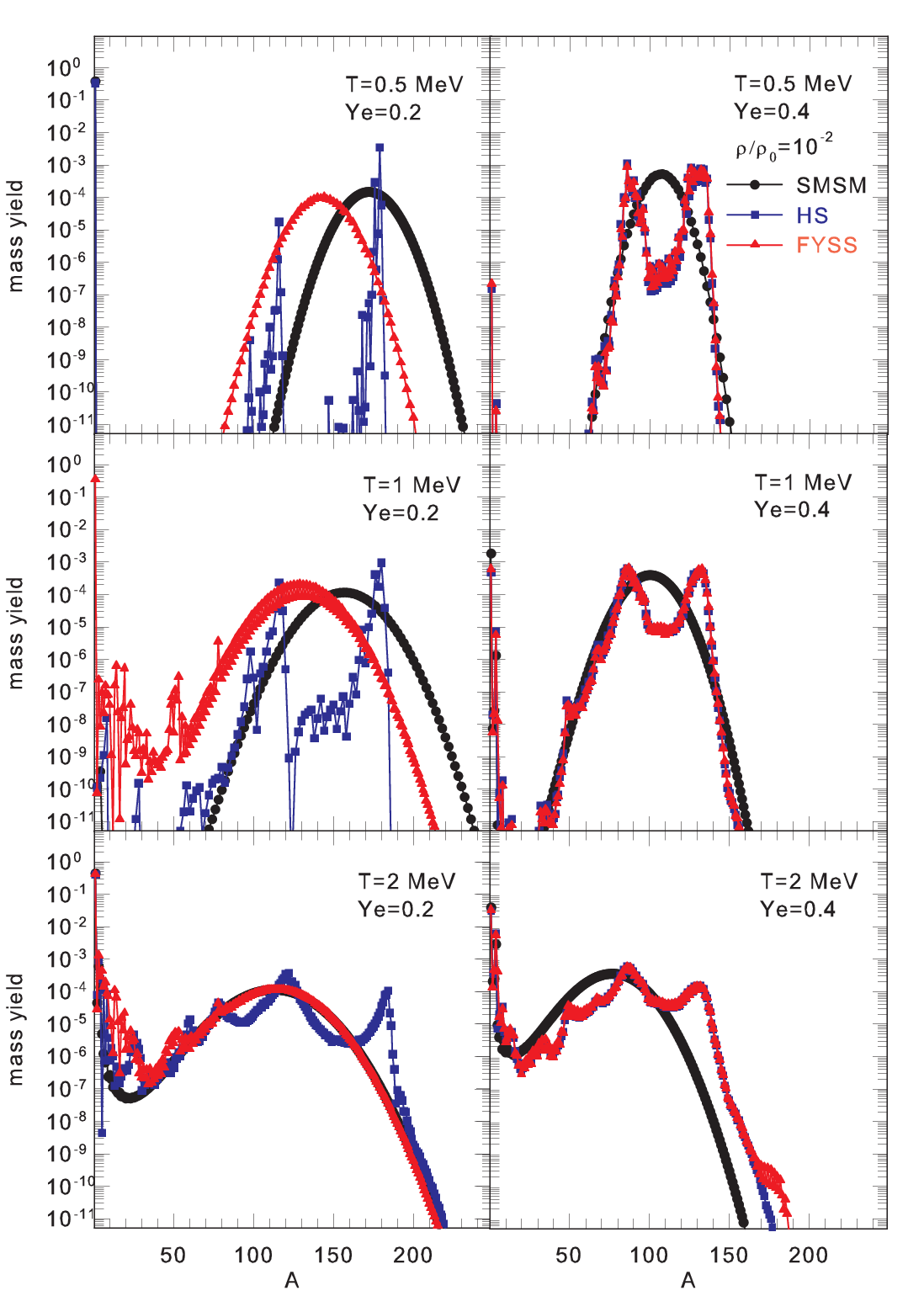}
\end{center}
\caption{\label{fig_ya3}\small{Mass distributions of fragments produced in matter with 
temperatures $T=0.5, 1$ and 2 MeV, electron
fractions $Y_e=0.2$ and 0.4, and  density $\rho/\rho_0=10^{-2}$. (Color version online.)
 }}
\end{figure}

\begin{figure}  
\begin{center}
\includegraphics[width=12cm,height=16cm]{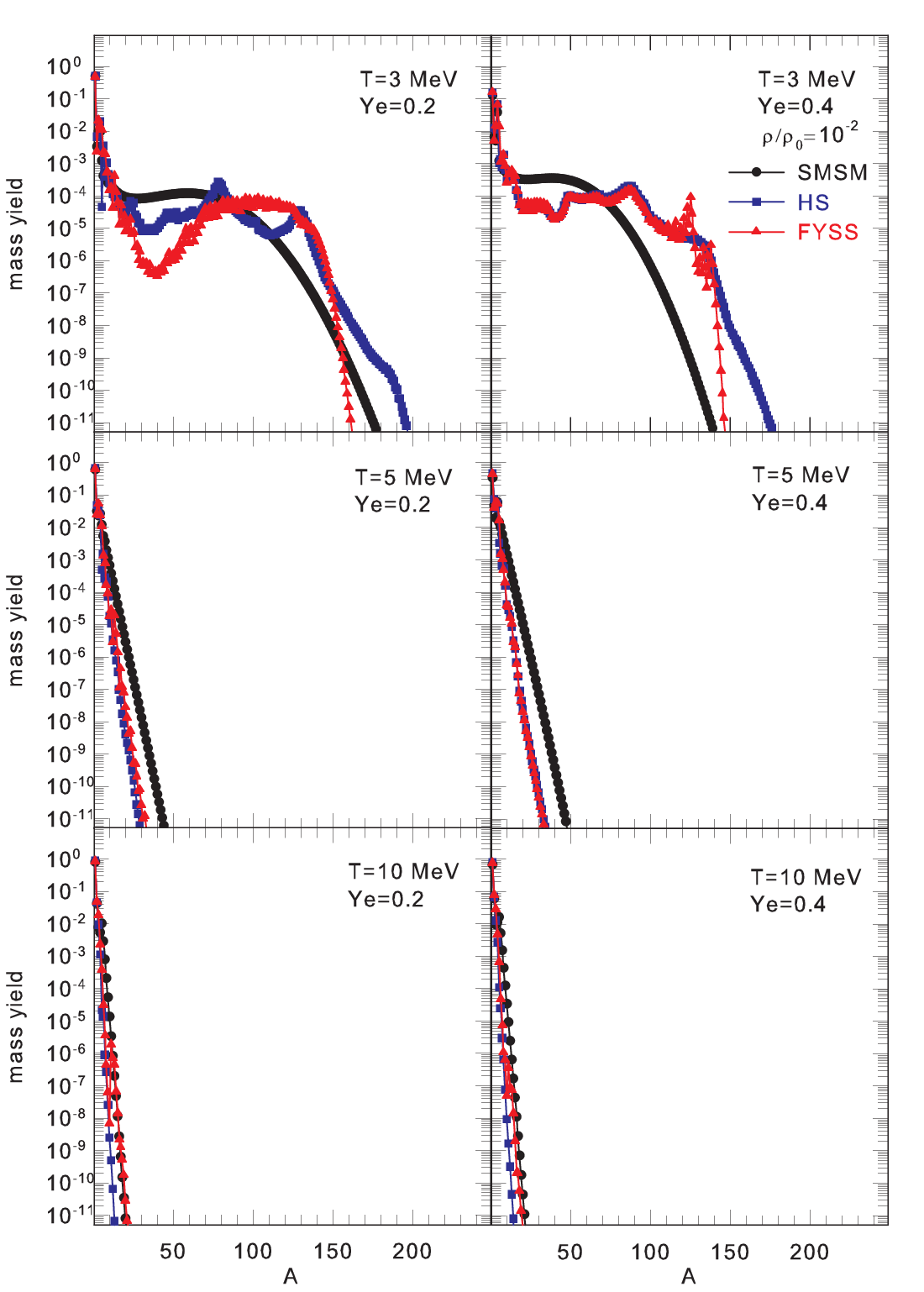}
\end{center}
\caption{\label{fig_ya4}\small{Mass distributions of fragments produced in matter with 
temperatures $T=3, 5$ and 10 MeV, electron
fractions $Y_e=0.2$ and 0.4, and  density $\rho/\rho_0=10^{-2}$. (Color version online.)
 }}
\end{figure}

\begin{figure}  
\begin{center}
\includegraphics[width=12cm,height=16cm]{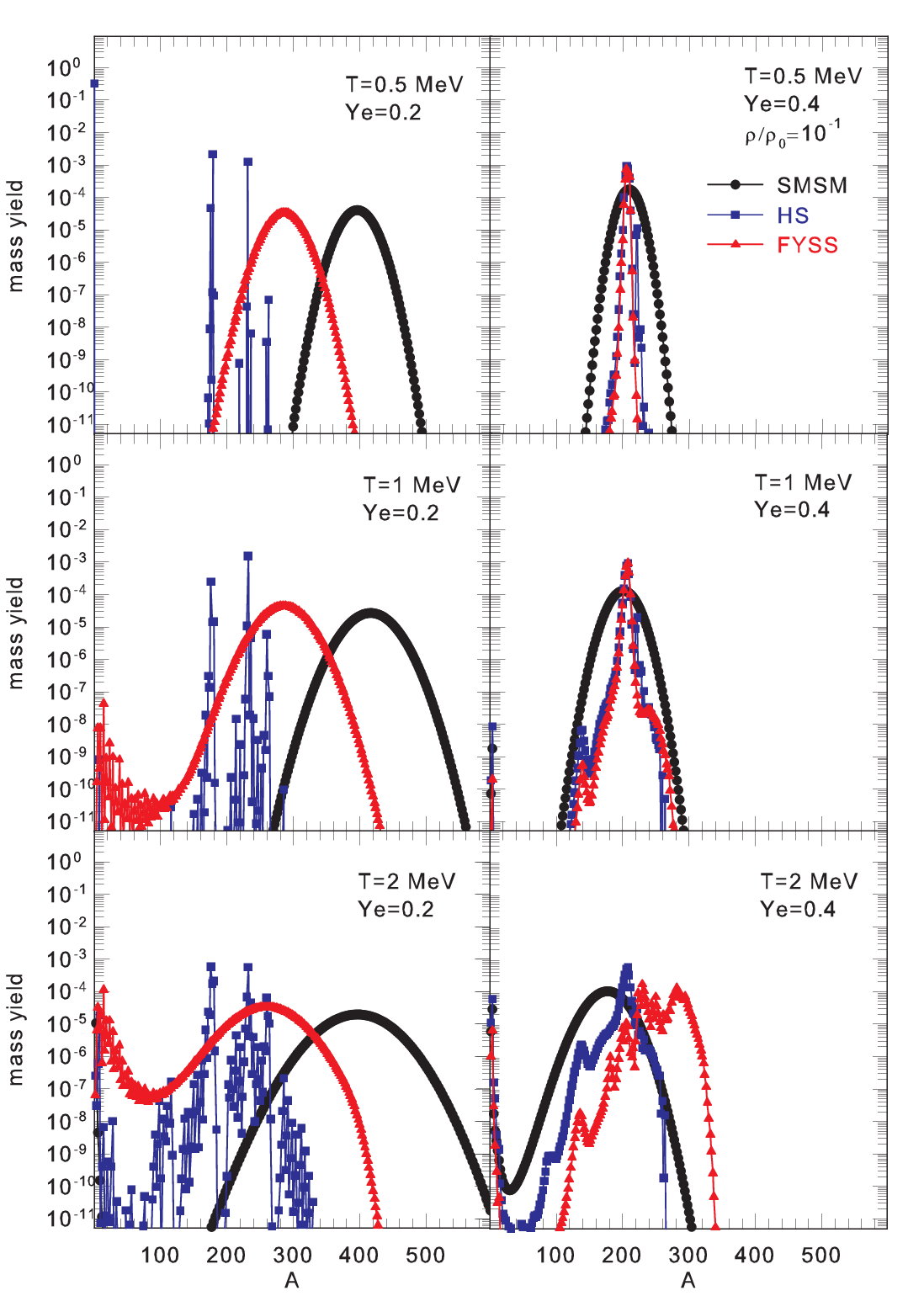}
\end{center}
\caption{\label{fig_ya5}\small{Mass distributions of fragments produced in matter with 
temperatures $T=0.5, 1$ and 2 MeV, electron fractions $Y_e=0.2$ and 0.4, 
and  density $\rho/\rho_0=10^{-1}$. (Color version online.)}}
\end{figure}

\begin{figure}  
\begin{center}
\includegraphics[width=12cm,height=16cm]{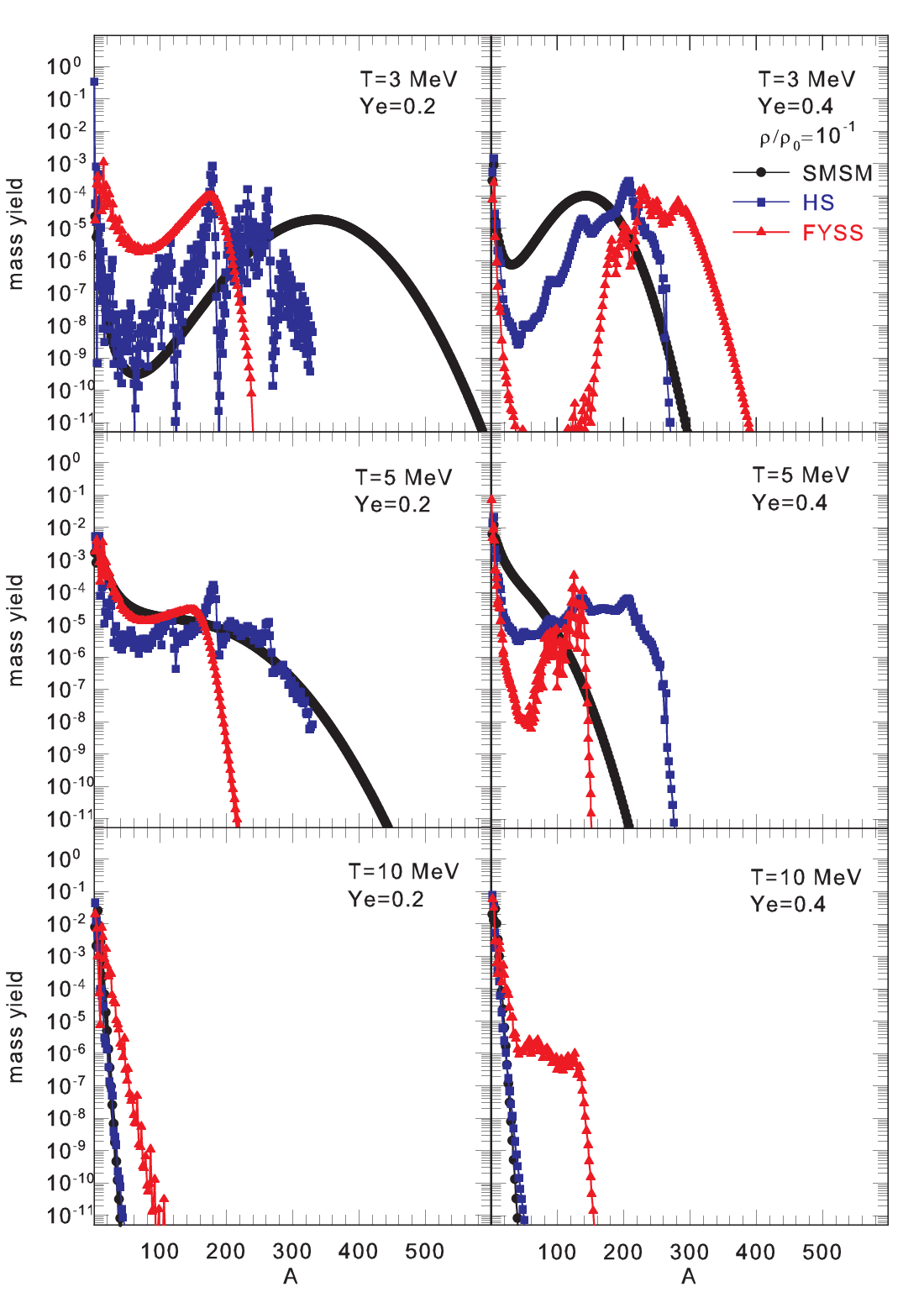}
\end{center}
\caption{\label{fig_ya6}\small{Mass distributions of fragments produced in matter with 
temperatures
$T=3, 5$ and 10 MeV, electron fractions $Y_e=0.2$ and 0.4, and  density 
$\rho/\rho_0=10^{-1}$. (Color version online.)
 }}
\end{figure}

Let us start with the analysis of mass distribution of nuclear species 
produced in stellar matter. 
Detailed comparison of mass distributions predicted by our models 
is presented in Figs.~\ref{fig_ya1}--\ref{fig_ya6}. It includes yields at 
various densities 
$\rho$, electron fraction $Y_e$ and 
temperature $T$. These distributions contain important information about 
fragmentation of matter in the nuclear liquid-gas phase transition 
(coexistence region). The concept of statistical equilibrium assumes a 
continuous interaction between fragments via specific microscopic 
processes, like absorption and emission of neutrons, which provide 
equilibration (see, e.g., discussion in Ref.~\cite{Botvina10}). In this 
situation the nuclei can remain hot and have modified properties and masses, 
which are different from the cold isolated nuclei. 

Figs.~\ref{fig_ya1} and \ref{fig_ya2} present results obtained at the 
lowest density under 
investigation $\rho/\rho_0=10^{-3}$. In this case the residual 
interaction between nuclear species should be minimal, though in-medium 
mass modifications are still possible. 
At very low temperatures the SMSM predicts a Gaussian-like distribution for 
heavy nuclei. In this case an approximation of a single heavy 
nucleus adopted in the Lattimer and Swesty \cite{Lattimer} and 
Shen et al. \cite{Shen} EOSs may work reasonably well for 
calculations of thermodynamical characteristics of matter. However, 
already at $T \goo 1$ MeV the gap between the Gaussian peak and light 
clusters and nucleons is essentially filled by nuclei of intermediate masses 
leading to characteristic U-shaped distributions. These distributions
are 
also typical for the onset of the liquid-gas phase transition in
finite nuclei 
\cite{SMM}. On the other side of the Gaussian peak for
big fragments, 
there is an continuous exponential fall of fragment yields with $A$. 
The difference from a single nucleus case is even more 
evident if we include nuclear shells, as done in HS and FYSS models. 

We remind that shell effects in masses 
of cold isolated nuclei may survive in nuclei at 
low temperatures ($T<1-2$ MeV) and low densities of matter. 
This can be seen in the distributions of HS and FYSS in Fig.~\ref{fig_ya1}. 
The peaks of the distributions occur around the well-known neutron magic 
numbers due to the increased binding energies. 
For most conditions the results of HS and FYSS look similar to the Gaussian 
distributions of SMSM with additional peaks on top. Only for $T=0.5$ and 1~MeV 
with $Y_e=0.2$ some interesting features occur. 
For $T=1$~MeV and $Y_e=0.2$, the yields of FYSS are similar to those of 
HS up to $A\sim 90$. In the intermediate mass range $90 < A < 130$ the 
yields of HS are several orders of magnitude larger. These are the neutron-rich nuclei 
contained in the theoretical nuclear structure calculations of 
Geng et al. \cite{Geng05}, but which are not in the experimental 
compilation of Audi et al.~\cite{Audi}. The jump around $A\sim130$ in 
FYSS is caused by the transition to the liquid-drop formulation for exotic 
nuclei. The differences visible for $T=0.5$~MeV and $Y_e=0.2$ can be 
explained in the same way. For these conditions, the HS model mainly 
contains nuclei with binding energies from the
theoretical nuclear structure calculation, which are absent in FYSS. 
Because the nuclei found in FYSS are not yet described by the liquid-drop 
model, this leads to a sharp peak for a certain nucleus with experimentally 
measured binding energy. The drop in HS around $A\sim120$ occurs because 
extremely neutron-rich nuclei with large
mass 
numbers are not included. This drop does not occur for $T=1$~MeV,
because the temperature 
increases the unbound neutron mass fraction and thereby decreases the 
asymmetry of heavy nuclei. For the larger values of $Y_e=0.4$ in 
Fig.~\ref{fig_ya1} no unexpected features occur, because nuclei have a 
smaller asymmetry and they are well inside the region of nuclei with 
measured binding energies. 

At temperature $T=2$ MeV we obtain plateau-like mass distributions in all 
three models. They are well known from nuclear multifragmentation studies 
\cite{SMM} and can be connected to the liquid-gas phase transition. 
Furthermore, it is impossible to 
describe it with a single nucleus approximation. As shown in 
Ref.~\cite{Botvina10} this phase transition can be driven by 
both temperature and density. Besides changing mass distributions 
qualitatively, at this point one can observe other critical phenomena as 
maximum heat capacity (i.e., a plateau-like behavior of the caloric 
curve), a minimum of exponent $\tau$  
in the power-law mass distribution $A^{-\tau}$ 
of intermediate mass fragments, large fluctuations in fragment 
sizes, etc., which exist in both finite and infinite systems 
(see, e.g., discussions in Refs.~\cite{Dag,Das05,Bugaev01}). 

At high temperatures ($T\geq 3$ MeV) all models predict disintegration of 
nuclear matter into small fragments and their yields decrease exponentially 
with mass number $A$. These results demonstrate that the transition 
from heavy nuclei (droplets of nuclear liquid) to lightest fragments and 
nucleons (nuclear gas) always proceeds through the same sequence of mass 
distributions: U-shape, power-law, and exponential fall-off, both with 
increasing temperature and decreasing density. Presence of mass shell 
effects in nuclei and other differences in their description do not 
influence this general evolution. For example, larger yields of heavy 
fragments in SMSM shown in Fig.~\ref{fig_ya2} can be explained by a difference 
from other models in the calculation of binding energies and
of the internal
excitation energy of 
intermediate-mass fragments leading to their higher entropy. For $A \leq 4$, where
all the models use almost the same binding energies, the yields are rather similar.

As was mentioned, at larger densities the mean-field 
effects start to play an important role in the HS and FYSS cases, 
and this influences the yields of nuclei. 
However, as we see from Figs.~\ref{fig_ya3}--\ref{fig_ya6} 
the general trends for the mass distributions do not change. 
It is natural that at high density heavy nuclei can be produced, 
up to $A\sim 500$. The temperature associated with the plateau-like 
behavior of mass distributions increases to $\approx 3$ MeV 
at $\rho/\rho_0=10^{-2}$, and to 4--6 MeV at $\rho/\rho_0=10^{-1}$ 
(as expected, at large $\rho$ it becomes more sensitive to the model). 
The temperature observed experimentally in
multifragmentation 
reactions is around $T\approx 5$ MeV \cite{Pochodzalla}. 

Apart from the overall similar behavior some interesting new features can 
be noticed in Figs.~\ref{fig_ya3}--\ref{fig_ya6}. In the upper and middle 
right panels of Fig.~\ref{fig_ya3} HS and FYSS give distributions with two 
separate peaks, corresponding to nuclei with magic neutron numbers 50 and 
82. Obviously such a bimodal distribution cannot be captured by the average 
value or by a single representative heavy nucleus. The upper 
left panel of Fig.~\ref{fig_ya3} shows the mass yields for very asymmetric, 
cold nuclear matter. Compared to the lower density shown in the upper left 
panel of Fig.~\ref{fig_ya1}, the neutron yield has further increased and 
neutrons start to become degenerate. This would occur e.g.~in the crust of 
a neutron star. In HS, shell effects are still dominant, leading to even 
sharper peaks than in Fig.~\ref{fig_ya1}, because the decreased role of 
temperature at higher density. Furthermore, for such high neutron 
abundances there are 
only few nuclei in the mass tables of HS with a suitable asymmetry. Note 
that the most abundant nucleus is at a similar $A$ for the maximum 
yield of the SMSM. Contrary to HS, the nuclei of FYSS are described by the 
liquid-drop formula, leading to Gaussian distributions like in the SMSM, 
but with smaller masses due to reduced surface energy. 

It is interesting to investigate differences between calculations with 
electron fractions $Y_e$=0.2 and 0.4. The calculations with large electron  fraction (respectively total proton fraction) are
more reliable, since nuclei with 
the corresponding $Z/A$ ratios have been studied in experiments. However, 
because of the electron capture at subnuclear densities, nuclei with low 
proton fractions may occur in supernova matter too. 
Regarding core-collapse supernovae, heavy nuclei are most important for the 
collapse phase before bounce. Here one has typically moderate asymmetries 
of $0.2 \loo Y_e \loo 0.4$, see e.g.~Refs.~\cite{Janka1,Sumi,Hempel11}. 
After the shock has formed, the temperature increases significantly and 
heavy nuclei are dissociated, as we also find in our comparison. In the 
late post bounce phase there appear regions with low $Y_e\sim 0.1$ in the 
shock-heated matter, but then the mass fraction of heavy nuclei is rather 
small, around 10$^{-3}$ to 10$^{-2}$. Therefore neutrino reactions with 
unbound neutrons and protons are more important. In the later stages of the 
protoneutron stars' cooling large asymmetries and moderate temperatures 
can be realized together, when the system approaches the equilibrium 
configuration of the neutron stars' inner crust. At a small 
electron fraction the SMSM can give very large and very neutron-rich nuclei 
which are not present in HS and FYSS tables. Due to the limitation of mass tables, one can see 
clear cuts of mass distributions in the HS calculations. 
The SMSM considers the whole ensemble of nuclei produced in stellar 
matter, without any additional constraint on their masses and charges. 
The universal liquid-drop approximation is used to describe their properties, 
and this is the reason why all SMSM distributions are smooth. 
The FYSS model also considers all possible exotic nuclei and they are calculated by a liquid-drop description when the mass or charge number 
of a nucleus is not contained in the table. As a result 
of the transition to the liquid-drop description some FYSS distributions 
becomes also smooth at low $Y_e$. 
It is also interesting that the mass numbers of the most abundant nuclei 
are different between the SMSM and FYSS model for $Y_e = 0.2$ as shown in 
Fig~\ref{fig_ya5} and Fig~\ref{fig_ya6}. This results from the difference 
of the two liquid-drop descriptions. The surface tensions of nuclei of SMSM 
depend on temperature but do not depend on neutron-richness of nuclei. 
On the other hands FYSS assumed the surface tension depending on 
neutron-richness but independent of temperature. For example,
the lower surface 
tensions of neutron-rich nuclei  of FYSS increase abundances of lighter 
nuclei in neutron-rich environment with $Y_e=0.2$, as shown in the left panel 
of Fig~\ref{fig_ya5}. 

We have already mentioned that the problem of the adequate description of nuclei under these extreme conditions, leading to modified binding energies and other properties, should be addressed in future studies. In this work we have taken into account only some of them which were generally discussed previously: the screening of Coulomb energy is calculated in the Wigner-Seitz approximation is considered in all three models, the temperature dependence of bulk and surface energies, as well as disappearance of shell effects at high excitations are taken into account in SMSM, the density dependence of the bulk and surface energies is included in the FYSS. Other modifications are now under intensive theoretical and experimental investigation in multifragmentation reactions, for example, a decrease of the symmetry energy of fragments \cite{LeFevre, Souliotis,Ogul}, and reduction of the binding energies of light clusters \cite{typel10,roepke11,Hagel} in-medium. A calculation within a self-consistent microscopic approach has been performed recently in \cite{de2012}. Presently, we can only declare significant differences between mass distributions of SMSM, HS and FYSS models, which use different assumptions on properties of fragments, especially at low electron fractions, low temperatures and high densities. We note that these differences are also manifested in the behavior of nuclear pressure, see Fig.~\ref{fig_pressure}(a)-(b). 

\vspace{0.3cm}
\subsection{Mass fractions}
\vspace{0.3cm}

\begin{figure}  
\begin{center}
\includegraphics[width=12cm,height=14cm]{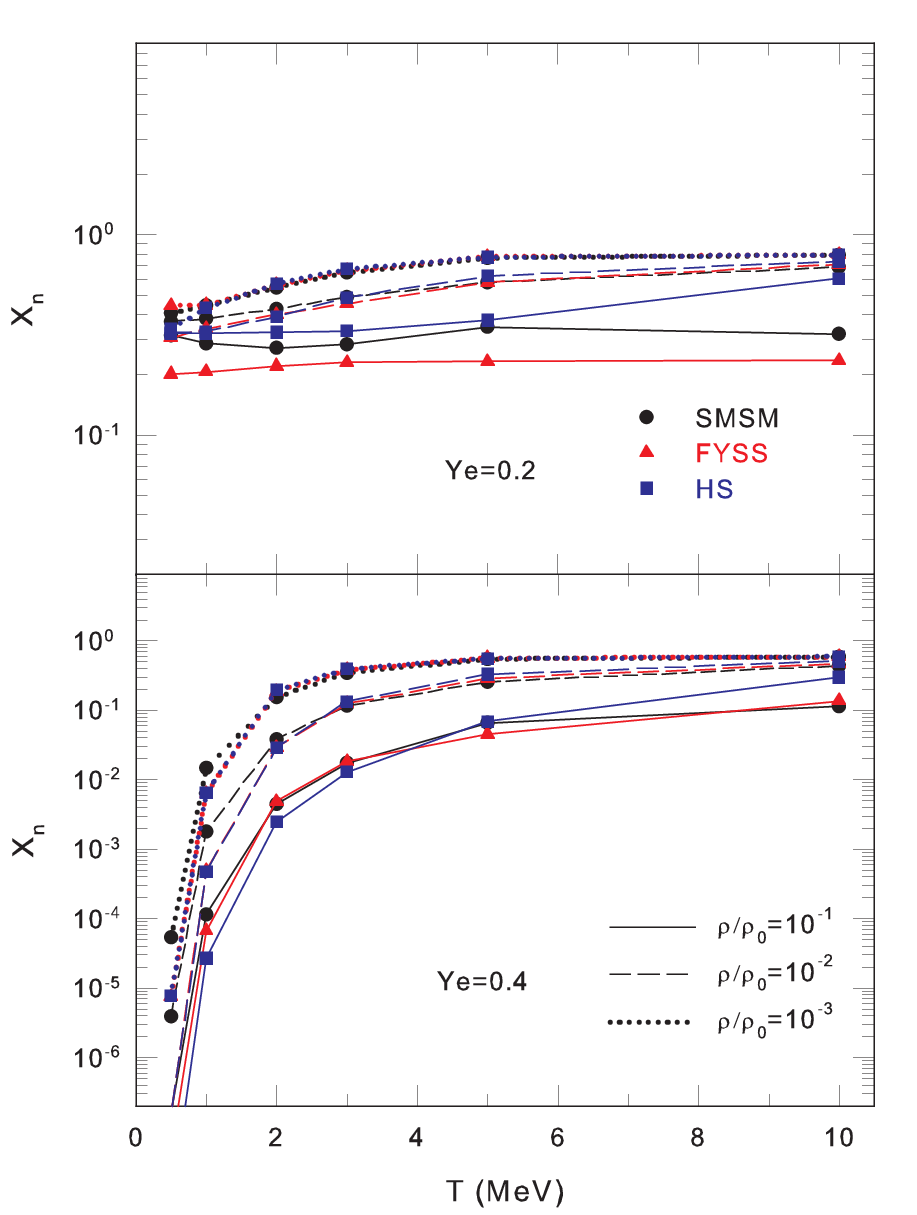}
\end{center}
\caption{\label{fig_xn}\small{Comparison of SMSM, HS, and FYSS model results for the average fraction of free neutrons as a function of temperature.  (Color version online.)
 }}
\end{figure}

\begin{figure}  
\begin{center}
\includegraphics[width=12cm,height=14cm]{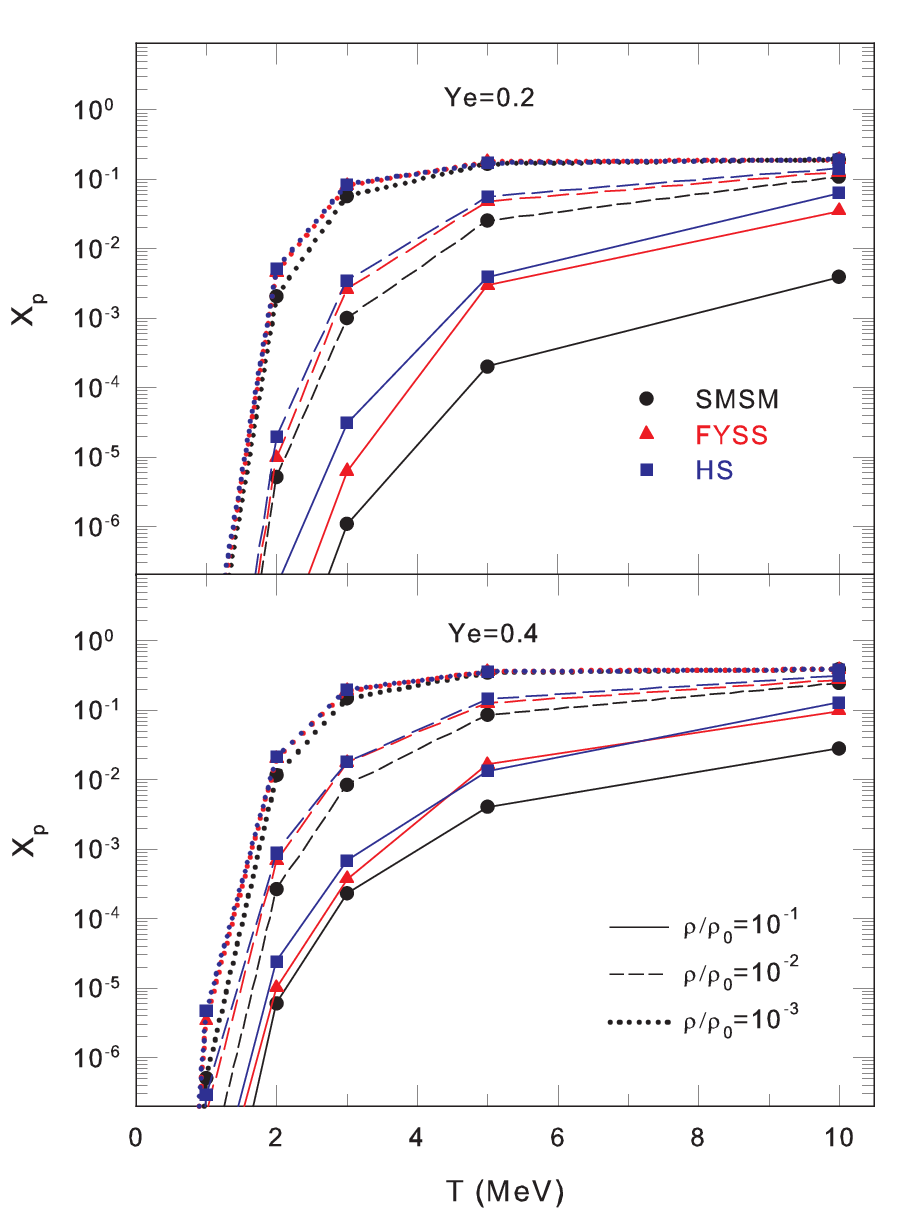}
\end{center}
\caption{\label{fig_xp}\small{Average fraction of protons as a function of temperature. (Color version online.)
 }}
\end{figure}

\begin{figure}  
\begin{center}
\includegraphics[width=12cm,height=16cm]{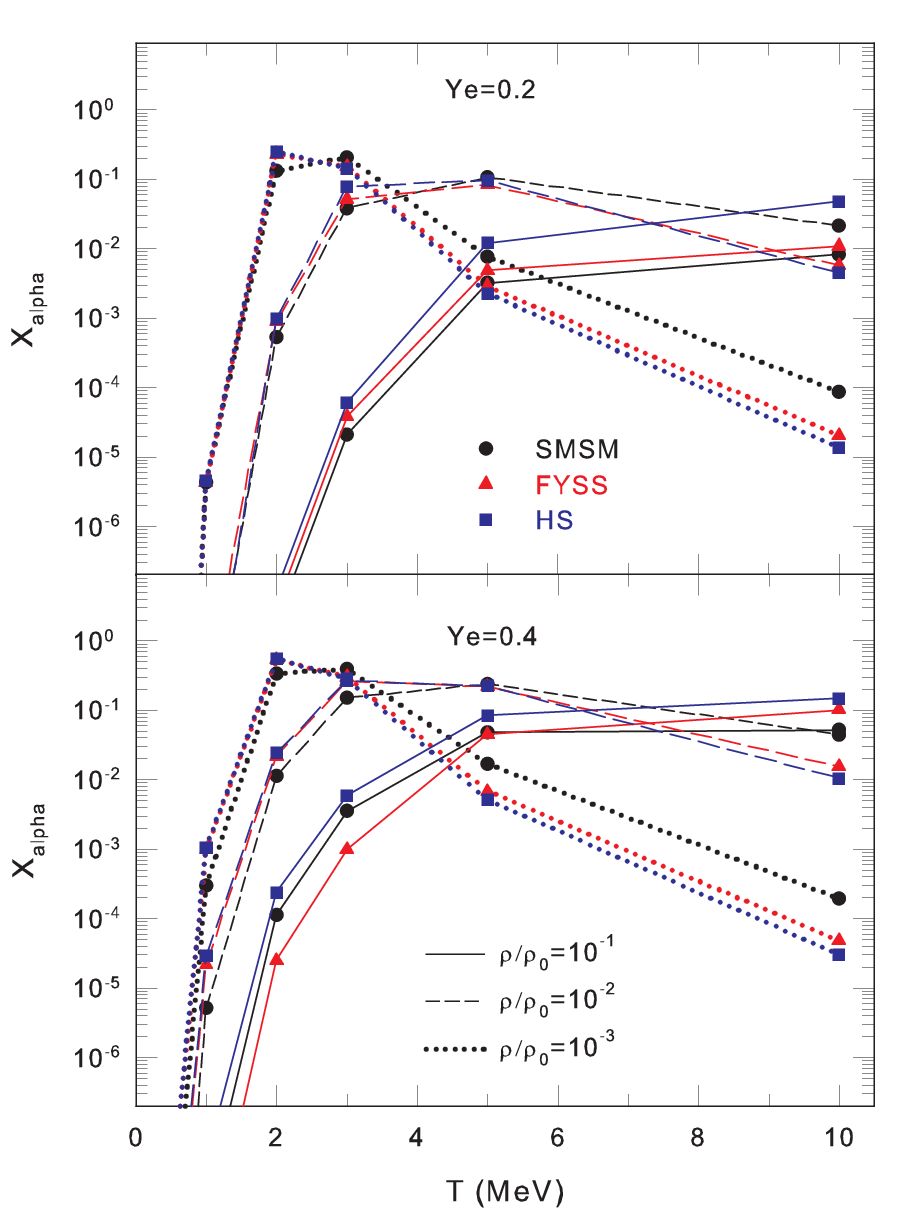}
\end{center}
\caption{\label{fig_xalpha}\small{Average fraction of alpha particles as a function of temperature. (Color version online.)
}}
\end{figure}

The mass fractions of light and heavy nuclei can be easily 
calculated from mass distributions presented above. The light nuclear 
species ($A \leq 4$) present the gas phase of 
nuclear matter, and they are mainly responsible for the nuclear pressure 
at high temperatures, together with unbound nucleons. In addition, many 
important reactions involve nucleons and alpha-particles, so it is 
crucial to know their abundances in stellar matter. 
We show the mass fractions of neutrons $X_n$, protons $X_p$, and alpha 
particles $X_{\alpha}$ in Figs.~\ref{fig_xn}, \ref{fig_xp}, and 
\ref{fig_xalpha}, respectively. 
All three models give similar results, except for $\rho/\rho_0=10^{-1}$ 
and $Y_e$=0.2 and at higher temperatures $T>5$~MeV and $Y_e=0.4$. 
In Fig.~\ref{fig_xn}, it is clear that the number of free neutrons 
decreases with 
increasing density, reflecting the formation of very heavy nuclei and the 
transition to the liquid phase at $\rho \rightarrow \rho_0$. The matter is 
mainly composed of heavy nuclei at low temperatures ($T<1$~MeV), however, 
the free neutrons are also present for small $Y_e=0.2$ 
(top panel of Fig.~\ref{fig_xn}), since the nuclear symmetry energy suppresses 
accumulation of neutrons in heavy nuclei. This can also be linked 
to the difference between the proton and neutron chemical potentials, 
which is much larger for $Y_e=0.2$ than for $Y_e=0.4$, 
see Figs.~\ref{fig_mup} and \ref{fig_mun}.

With increasing temperature heavy nuclei gradually disintegrate into 
$\alpha$'s, neutrons and protons. 
For this reason in Fig.~\ref{fig_xalpha} one can see a so-called 
"rise-and-fall" 
behavior of $X_{\alpha}$, which occurs actually for both increasing 
temperature and decreasing density. 
By disintegration of nuclei one can also explain an increase of mass 
fractions of protons $X_p$, which should reach $Y_e$ values at very high 
temperatures (Fig.~\ref{fig_xp}). We remark that other light nuclei, like 
deuterons or tritons, can appear with large abundance. For $T\geq 5$~MeV 
and at sufficiently low densities 
these light nuclei can be even more abundant than alpha particles 
\cite{Botvina10,Hempel10,Hempel11,typel10,Hempel11b,Sumi08,Arcones08}. 
In the present study, to be consistent with other works, 
we only show the mass fraction of alpha particles. 
The information from mass fractions is complementary to the one from 
mass distributions. It shows that at low densities like 
$\rho/\rho_0=10^{-3}$ the disintegration of matter into light particles 
happens already at temperatures $T \approx 1-2$ MeV, while 
at subnuclear densities ($\rho \approx 10^{-1}\rho_0$) some heavy nuclei 
survive even at high temperatures, though they become very excited. 
We show the mass fraction of heavy nuclei $X_{heavy}$ ($A>4$) in 
Fig.~\ref{fig_xheavy}.
At higher temperatures all EOS models give essentially different results. 
It is important to note that these ``heavy'' nuclei at high temperatures are actually close to the lower bound $A=5$, i.e., they are light and
intermediate mass nuclei. At high temperature the fractions of heavy nuclei in SMSM are higher than in HS
and FYSS, since SMSM takes into account the temperature dependences of bulk and surface
energies of such nuclei. The FYSS and HS also include the contribution to the free energy
due to the internal excitations, but only in the bulk term. In SMSM the surface contribution is important and it increases for light nuclei.

\vspace{0.3cm}
\subsection{Moments of mass and charge distributions}
\vspace{0.3cm}

One can define a mean mass number of the heavy fragments 
$\langle A_h \rangle$ (taking into account only nuclei with $A>4$). 
Actually, in the case of a sharp Gaussian mass distribution, it can be used 
for characterization of the nuclear liquid phase. 
It is also important for comparison of our approaches with 
models assuming a single nucleus approximation. 
Souza et al. have estimated \cite{souza08} that the mass number of 
the single nucleus approximation is systematically over-predicted compared 
to the average of a nuclear distribution calculated with otherwise the same 
nuclear physics inputs. We have analyzed the $\langle A_h \rangle$ evolution 
with respect to different temperatures 
for various densities and electron fractions in Fig.~\ref{fig_aheavy}. 
As expected, these values decrease with T, and then, beyond some value, 
decrease rather slowly, and go to a nearly constant value. 
This suggests that the vaporization process becomes dominant around this 
point. It can be expected that all models should approach the limiting value 
of $\langle A_h \rangle=5$ with increasing temperature. For the temperatures 
shown in Fig.~\ref{fig_aheavy}, this happens in all models, but only for the 
two lower densities. Note that the mass fractions of these nuclei are 
significantly decreased at the same time, see Fig.~\ref{fig_xheavy}. 
For $\rho/\rho_0=0.1$, the average mass numbers remain larger than five 
for all models even at $T=10$~MeV. Here it is found that FYSS gives the 
largest $\langle A_h \rangle$ around 10. These differences are a result 
of the different temperature and mass dependence of the free energies of 
nuclei in the models, as was discussed previously: In SMSM this is done via the temperature dependence 
of the liquid-drop formula, and in HS via the temperature-dependent internal 
partition function. In FYSS also the temperature-dependent internal 
partition function is used, but without an integration cut for high 
excitation energies. Furthermore, at large densities when nuclei are 
affected from the liquid-drop modifications of FYSS, there is an additional 
temperature dependence of the bulk energy of nuclei, calculated from the RMF 
model. In Fig.~\ref{fig_zheavy} we show the average charge of heavy nuclei 
$\langle Z_h \rangle$, which shows a similar trend as Fig.~\ref{fig_aheavy}. 

The standard deviation of the average mass of heavy nuclei, 
$\sigma_{A_h}=\sqrt{\langle{A_h}^2\rangle-\langle A_h \rangle^{2}}$ is 
shown in Fig.~\ref{fig_dispersion}. 
This figure gives us valuable information about the character of 
the nuclear distributions. 
It is seen for higher densities that all three models have a maximum 
around $T=5$ MeV which exactly corresponds to the temperatures obtained in 
investigations of nuclear multifragmentation of finite nuclei. 
The values of $\sigma_{A_h}$ in this region vary between 50 and 
100 for different models. 
Interestingly, the mass fractions of heavy nuclei are still 
dominant for these conditions. This large spread in the mass distribution 
could be important for neutrino reactions in core-collapse supernova. 
For lower densities the maximum $\sigma_{A_h}$ values shift to lower 
temperatures, which correspond to plateau-like 
mass distributions discussed above. Again, all EOSs demonstrate a
similar 
behavior except for higher density $\rho/\rho_0=10^{-1}$. 

\vspace{0.3cm}
\subsection{Isotopic distributions}
\vspace{0.3cm}

To get further insight into characteristics of nuclear species 
we have investigated the isotopic yields for selected elements 
with charges $Z=8, 26$, and $50$ described by the three models, 
see Figs.~\ref{fig_yiso1}--\ref{fig_yiso6}. We have selected 
only the cases, where the yields of isotopes are larger than 
$\sim 10^{-12}$. The study of isotopic yields helps to understand 
the differences observed for summed  quantities like mass yields or 
mass fractions discussed before. Indeed we observe similar features, 
but now in more detail. Generally, the 
SMSM gives Gaussian type distributions, which are the consequence of the 
liquid-drop description of fragments. At some conditions HS and FYSS 
results are similar because they use the same tabulated binding energies. 
This can be seen e.g.\ for the oxygen isotopes shown in Fig.~\ref{fig_yiso1}. 
We remark that no theoretically calculated binding energies are included 
in HS for oxygen. Therefore the used tabulated binding energies in FYSS and 
HS are identical, and one finds almost identical isotope distributions. For 
the larger densities shown in Fig.~\ref{fig_yiso2}, differences between 
HS and FYSS emerge even for nuclei which are contained in the mass table. 
This happens because in FYSS the bulk energies of these nuclei experience 
the liquid-drop medium modifications. In the HS case there are cuts, since 
the tables contain only a limited number of nuclei. 
This can also be seen in Fig.~\ref{fig_yiso2}, e.g. in the lower left panel, 
where the maximum $A$ of oxygen isotopes considered in HS is 28. In the 
FYSS model, nuclei which are not in the mass table are described exclusively 
with the RMF liquid-drop model. This concerns all oxygen isotopes with 
$A>28$. Jumps occur at the boundaries between the tabulated nuclei and the 
nuclei which are described by the liquid-drop model. Comparing HS 
and FYSS, it is very interesting to realize that the yields of HS seem to 
connect relatively smoothly with the liquid-drop yields of FYSS for $A>28$. 
This could be a result of rearrangement effects to satisfy the baryon and charge conservations (\ref{rhotot})
in the nuclear distributions. 
Apart from the upper right panels in Figs.~\ref{fig_yiso1} and 
\ref{fig_yiso2}, the purely liquid-drop description of the SMSM gives 
significantly larger oxygen yields, with up to four orders of magnitude 
difference. In addition to the different description of medium effects, this 
also represents the differences in the nuclear binding energies. The nuclear 
structure effects included in HS and FYSS can be identified, e.g., by the 
even-odd staggering and the favored appearance of $^{16}$O clearly visible 
at lower temperature. 

The abundances of the nuclei with $Z=8$ of the FYSS model  are increasing 
with mass number in left panel of  Fig~\ref{fig_yiso2}. It is a very interesting effect related to RMF description of nuclear 
clusters and properties of nucleons in dilute matter. One can see that 
at these conditions neutrons are accumulated in clusters considerably more than 
in the SMSM case. The isotope distribution of oxygen with a peak at 
$A \approx 40$ can be realized as effective reduction of the symmetry 
energy for these nuclei (see, Eq.~(\ref{fsym})). Such a trend have been 
already reported in analysis of some experimental data on multifragmentation 
\cite{LeFevre,Souliotis,Ogul,Iglio,Hudan,Ogul09,Henzlova}. 

The isotopic yields of iron nuclei are shown in Figs.~\ref{fig_yiso3} and 
\ref{fig_yiso4}. In HS, the theoretical nuclear structure calculations extend 
up to $^{92}$Fe. In FYSS, tabulated binding energies from Ref.~\cite{Audi} 
are only available up to $^{72}$Fe, and isotopes with mass numbers 
larger than 72 are calculated by the liquid-drop formulas. This leads to 
jumps in isotope distributions. Such jumps are avoided in the HS model 
by using binding energies from theoretical nuclear structure calculations 
for exotic nuclei. However, it is found that the end of the 
mass table is reached, e.g., in the upper left panel of Fig.~\ref{fig_yiso4}. 
Compared with the SMSM, the shapes of the iron isotope distributions 
predicted in HS and FYSS are much more similar than for oxygen. 
The mean mass  numbers of isotopes in the three models are rather close 
to each other. 
However, in SMSM the isotope yields are several orders of magnitude larger 
than in FYSS or HS, that is related to the mass distribution discussed 
above. 

In Fig.~\ref{fig_yiso5} we show the isotope distributions for tin nuclei, 
which are qualitatively rather similar to the iron case. 
Similar results as in the previous figures are obtained. 
In Fig.~\ref{fig_yiso6} we show yields of iron and tin isotopes, 
for the most extreme conditions, namely $\rho/\rho_0 = 0.1$ and $T=5$~MeV. 
There are rather large differences for all three models.
For low $Y_e$, the limited range of mass numbers in HS is apparent. 
SMSM obtains overall larger yields, as before. Also the mean values 
predicted by the three models are slightly shifted. Large discontinuities 
are again found in the FYSS calculations because of utilizing different 
descriptions of fragments for different mass regions, as previously noted. 
The abundance curves in SMSM at low $Y_e$ are steeper than in FYSS since 
FYSS includes RMF calculations of nuclear binding energies leading to the difference
in their symmetry energies.

\vspace{0.3cm}
\subsection{Thermodynamical properties}
\vspace{0.3cm}

\begin{figure} 
\begin{center}
\includegraphics[width=12cm,height=16cm]{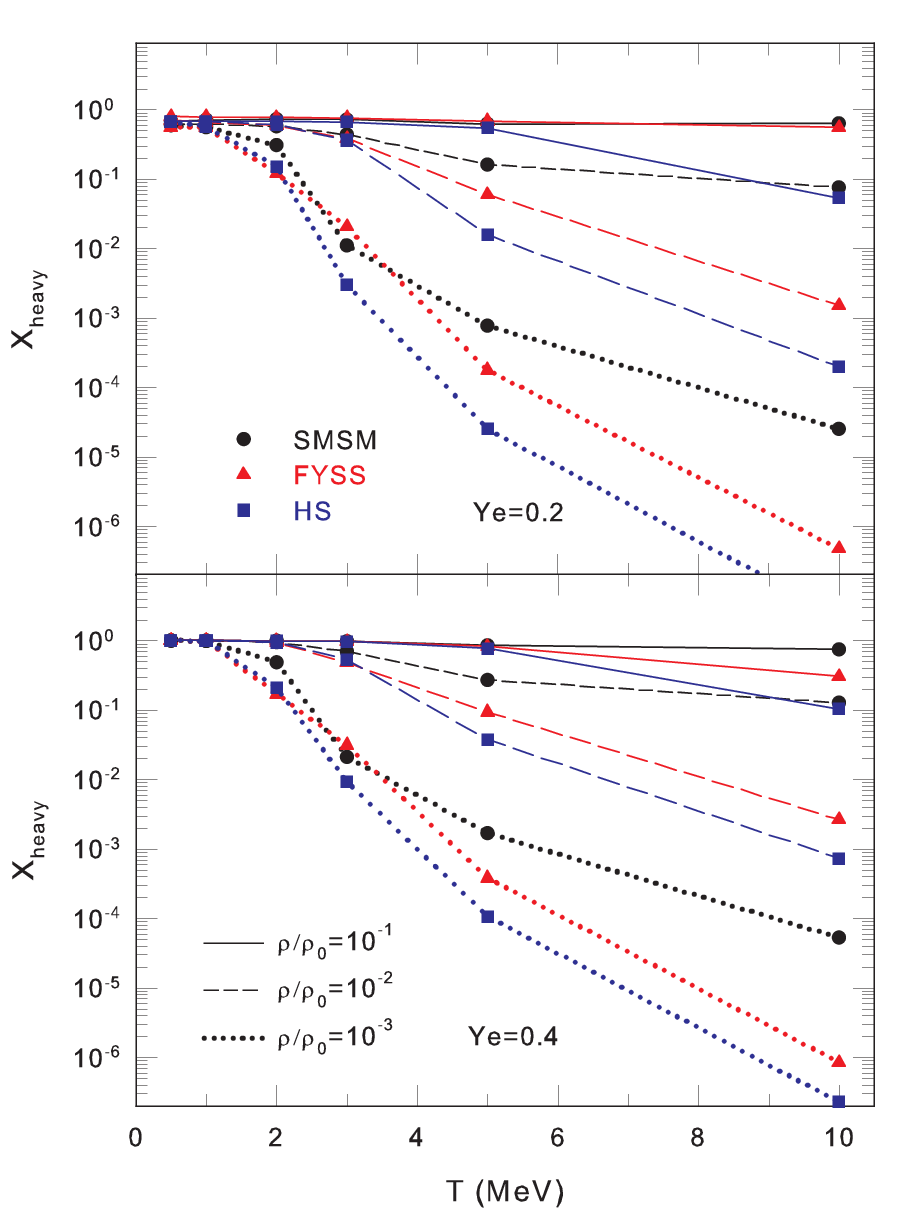} 
\end{center}
\caption{\label{fig_xheavy}\small{Average fraction of heavy particles ($A>4$) as a function of temperature. (Color version online.)
 }}
\end{figure}

\begin{figure}  
\begin{center}
\includegraphics[width=12cm,height=16cm]{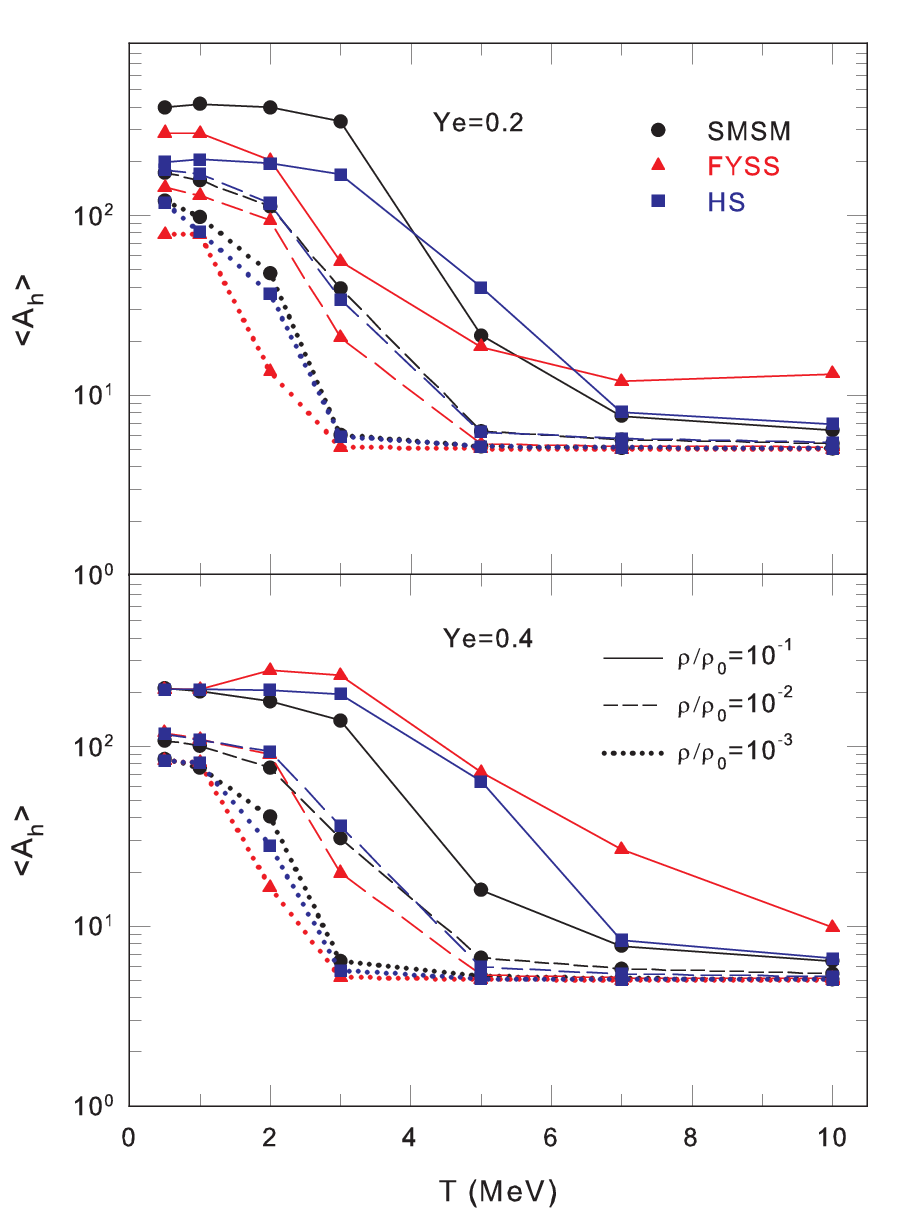}
\end{center}
\caption{\label{fig_aheavy}\small{Average mass number of heavy nuclei ($A>4$) as a function of temperature. (Color version online.)
 }}
\end{figure}

\begin{figure}  
\begin{center}
\includegraphics[width=12cm,height=16cm]{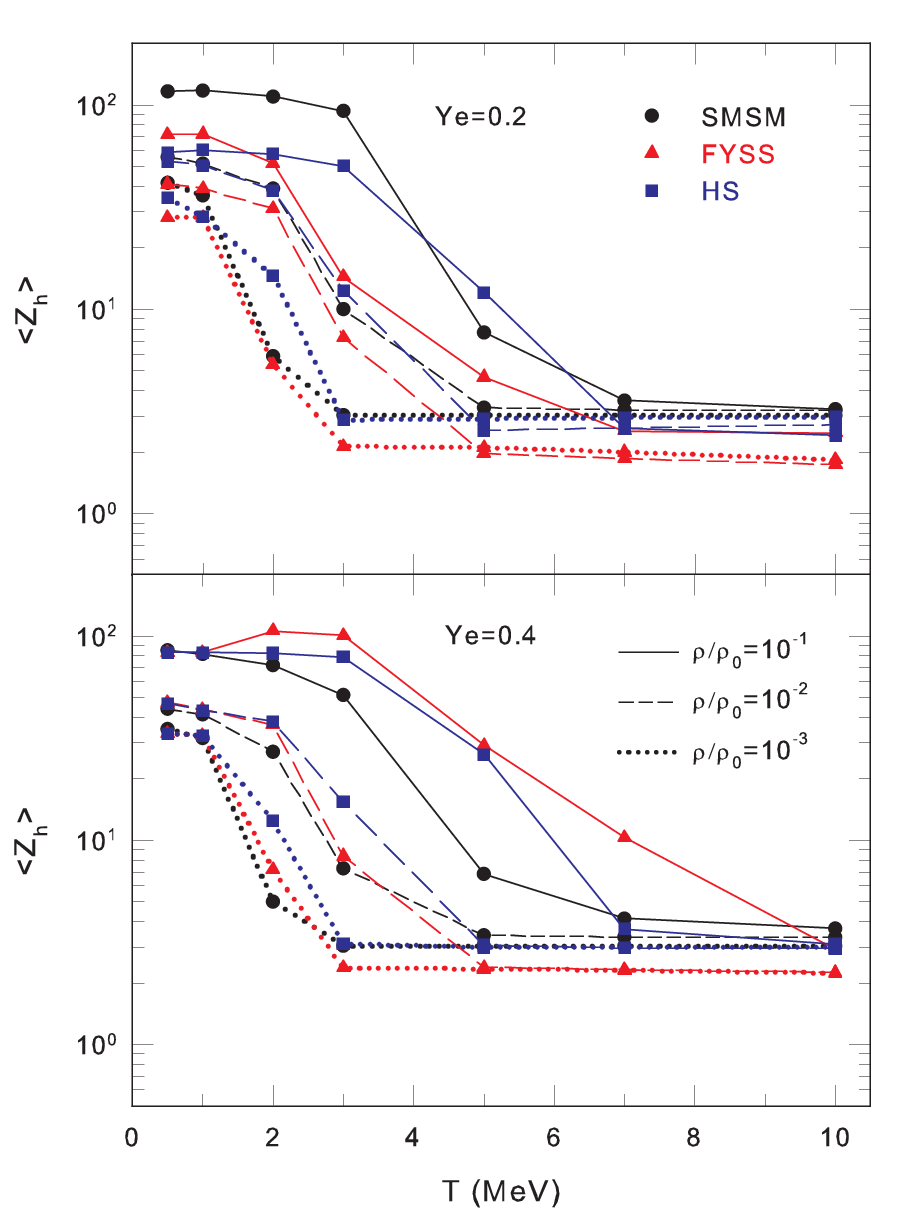}
\end{center}
\caption{\label{fig_zheavy}\small{Average proton number of heavy nuclei ($A>4$) as a function of temperature. (Color version online.)
 }}
\end{figure}

\begin{figure}  
\begin{center}
\includegraphics[width=12cm,height=16cm]{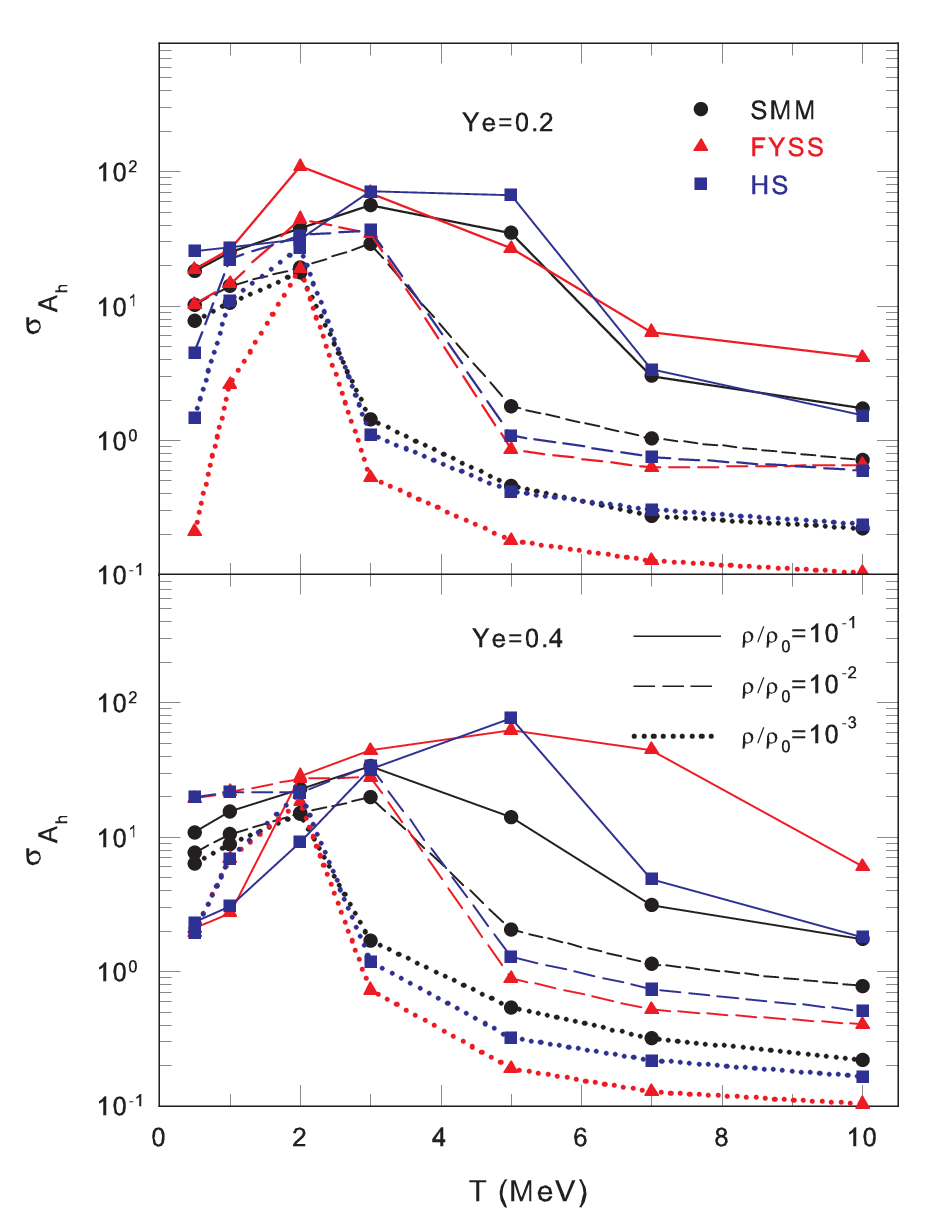}
\end{center}
\caption{\label{fig_dispersion}\small{Dispersion of the mass number of heavy nuclei ($A>4$) as a function of temperature. (Color version online.)
 }}
\end{figure}

\begin{figure}  
\begin{center}
\includegraphics[width=12cm,height=16cm]{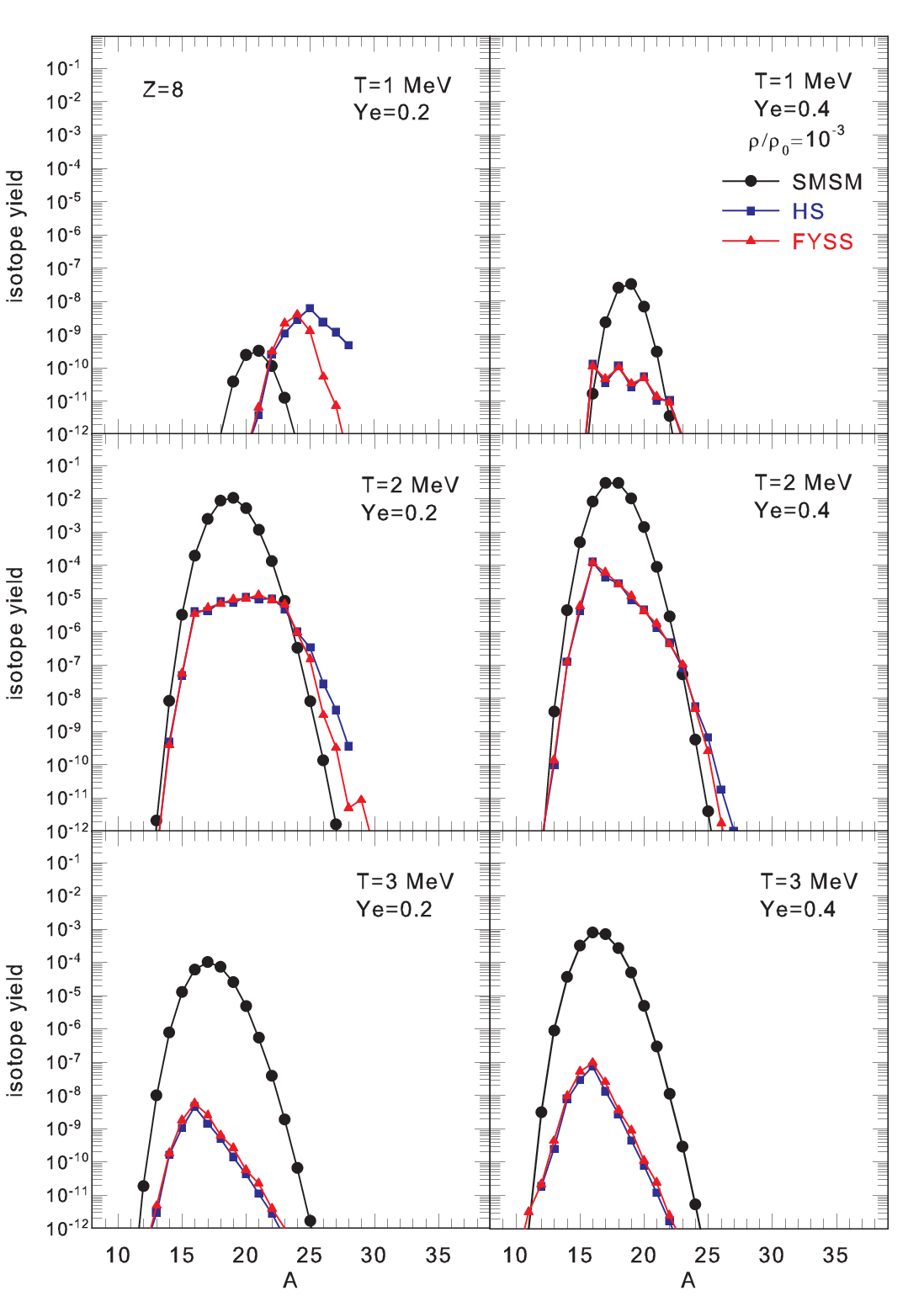}
\end{center}
\caption{\label{fig_yiso1}\small{Isotopic distributions of $Z=8$ fragments produced in matter with temperatures $T=1, 2$ and 3 MeV, electron
fractions $Y_e=0.2$ and 0.4, and  density $\rho/\rho_0=10^{-3}$. (Color version online.)
 }}
\end{figure}

\begin{figure}  
\begin{center}
\includegraphics[width=12cm,height=16cm]{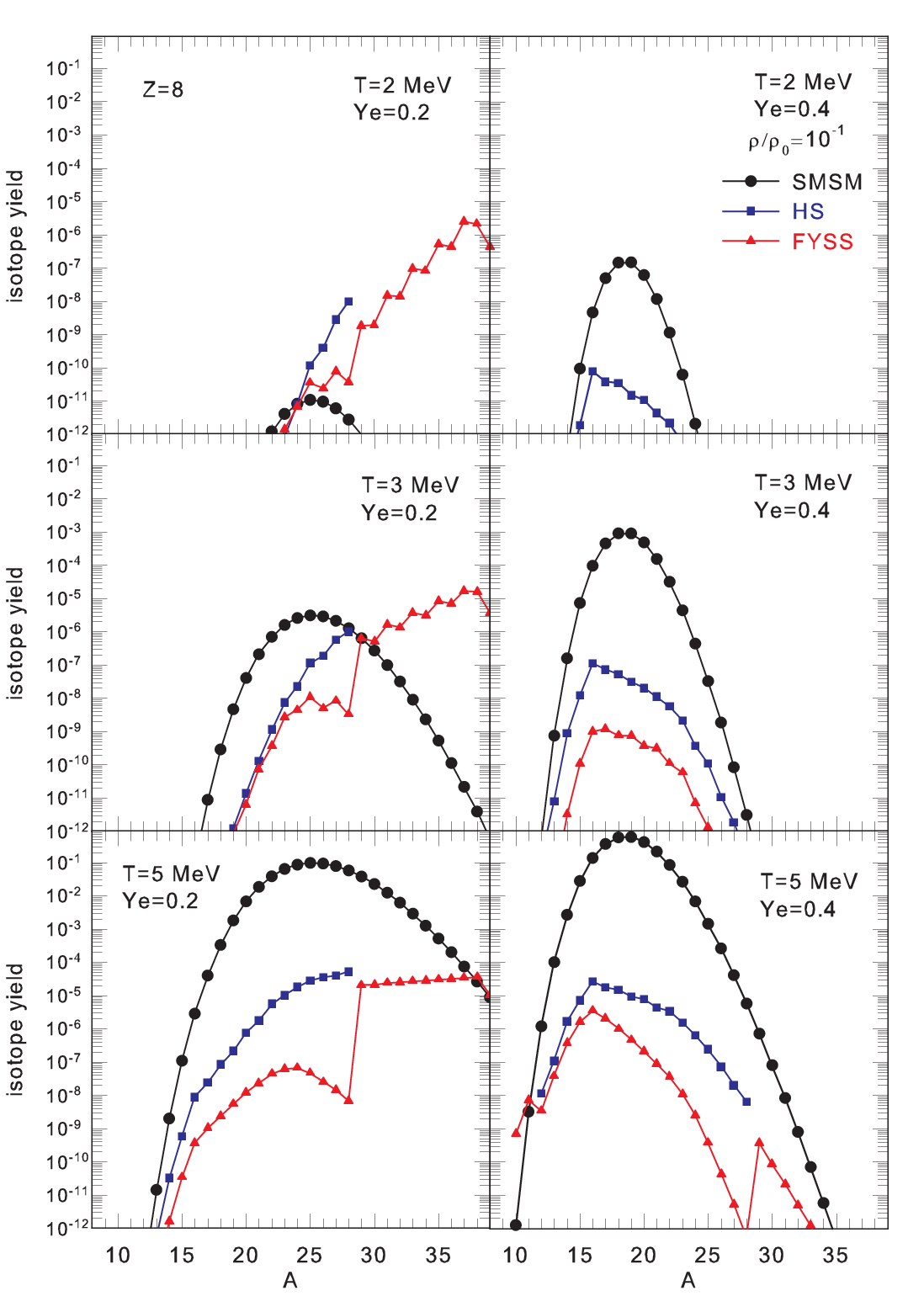}
\end{center}
\caption{\label{fig_yiso2}\small{Isotopic distributions of $Z=8$ fragments produced in matter with temperatures $T=2, 3$ and 5 MeV, electron
fractions $Y_e=0.2$ and 0.4, and  density $\rho/\rho_0=10^{-1}$. (Color version online.)
 }}
\end{figure}

\begin{figure}  
\begin{center}
\includegraphics[width=12cm,height=16cm]{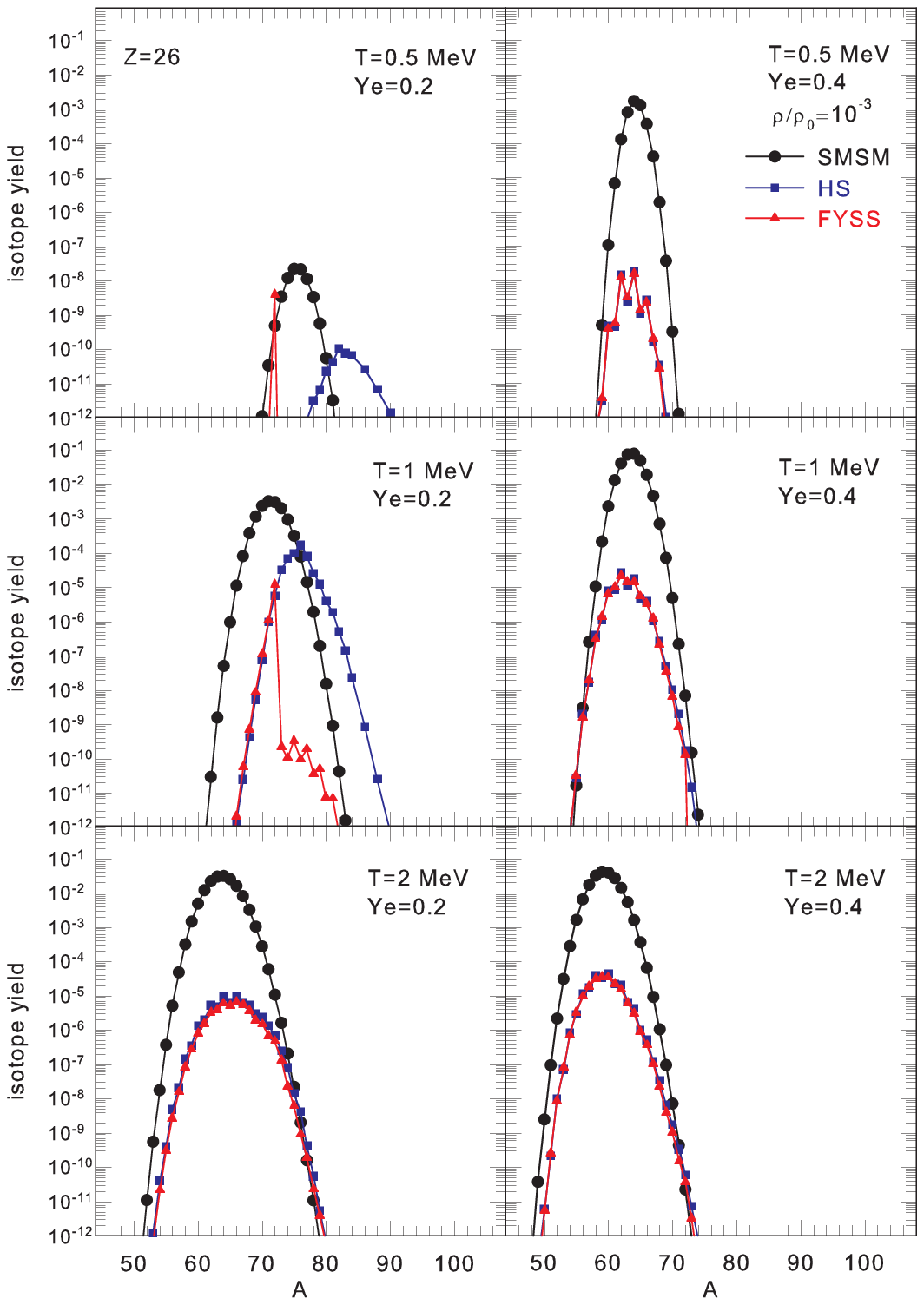}
\end{center}
\caption{\label{fig_yiso3}\small{Isotopic distributions of $Z=26$ fragments produced in matter with temperatures $T=0.5, 1$ and 2 MeV, electron
fractions $Y_e=0.2$ and 0.4, and  density $\rho/\rho_0=10^{-3}$. (Color version online.)
 }}
\end{figure}

\begin{figure}  
\begin{center}
\includegraphics[width=12cm,height=16cm]{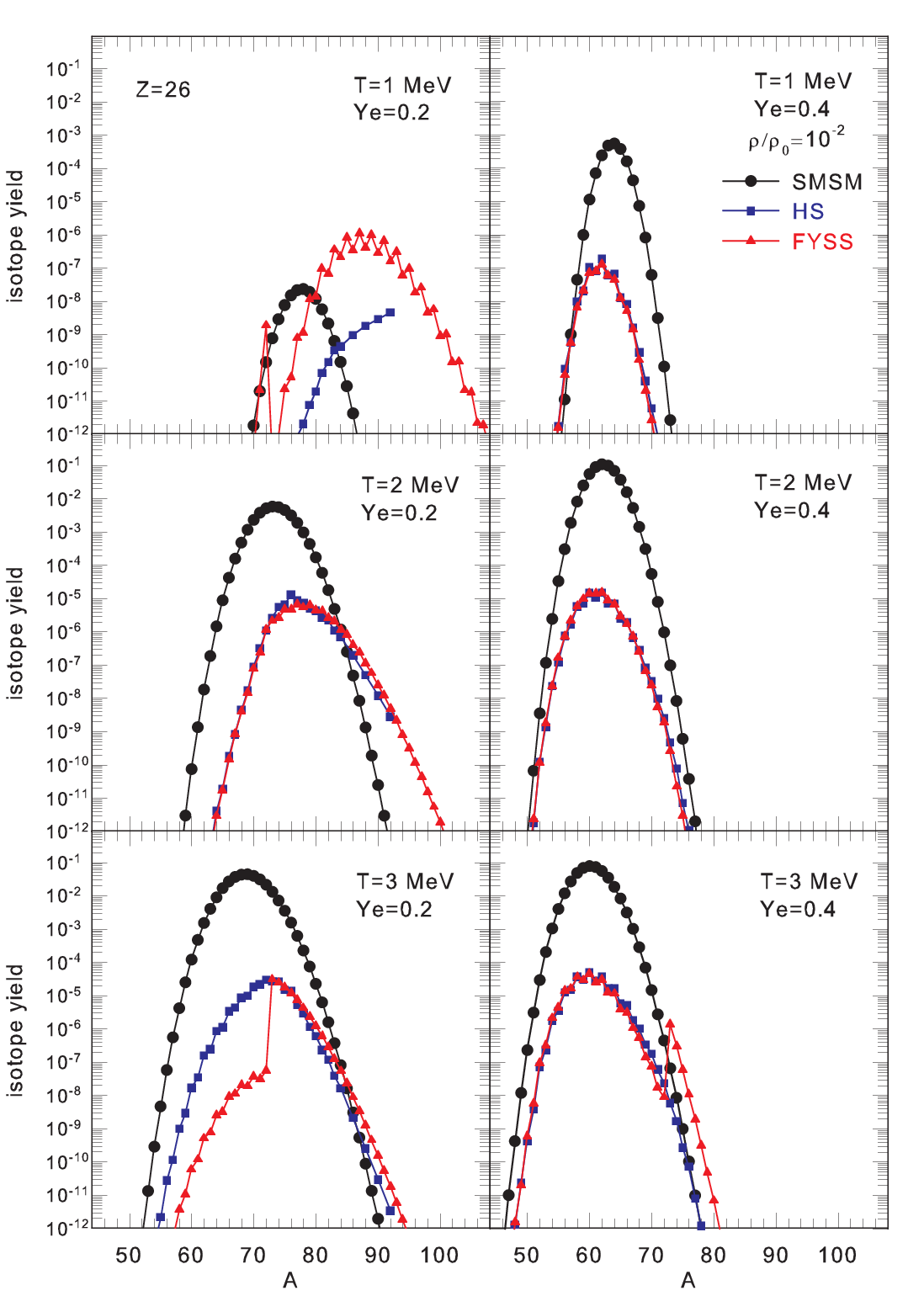}
\end{center}
\caption{\label{fig_yiso4}\small{Isotopic distributions of $Z=26$ fragments produced in matter with temperatures $T=1, 2$ and 3 MeV, electron
fractions $Y_e=0.2$ and 0.4, and  density $\rho/\rho_0=10^{-2}$. (Color version online.)
 }}
\end{figure}

\begin{figure}  
\begin{center}
\includegraphics[width=12cm,height=16cm]{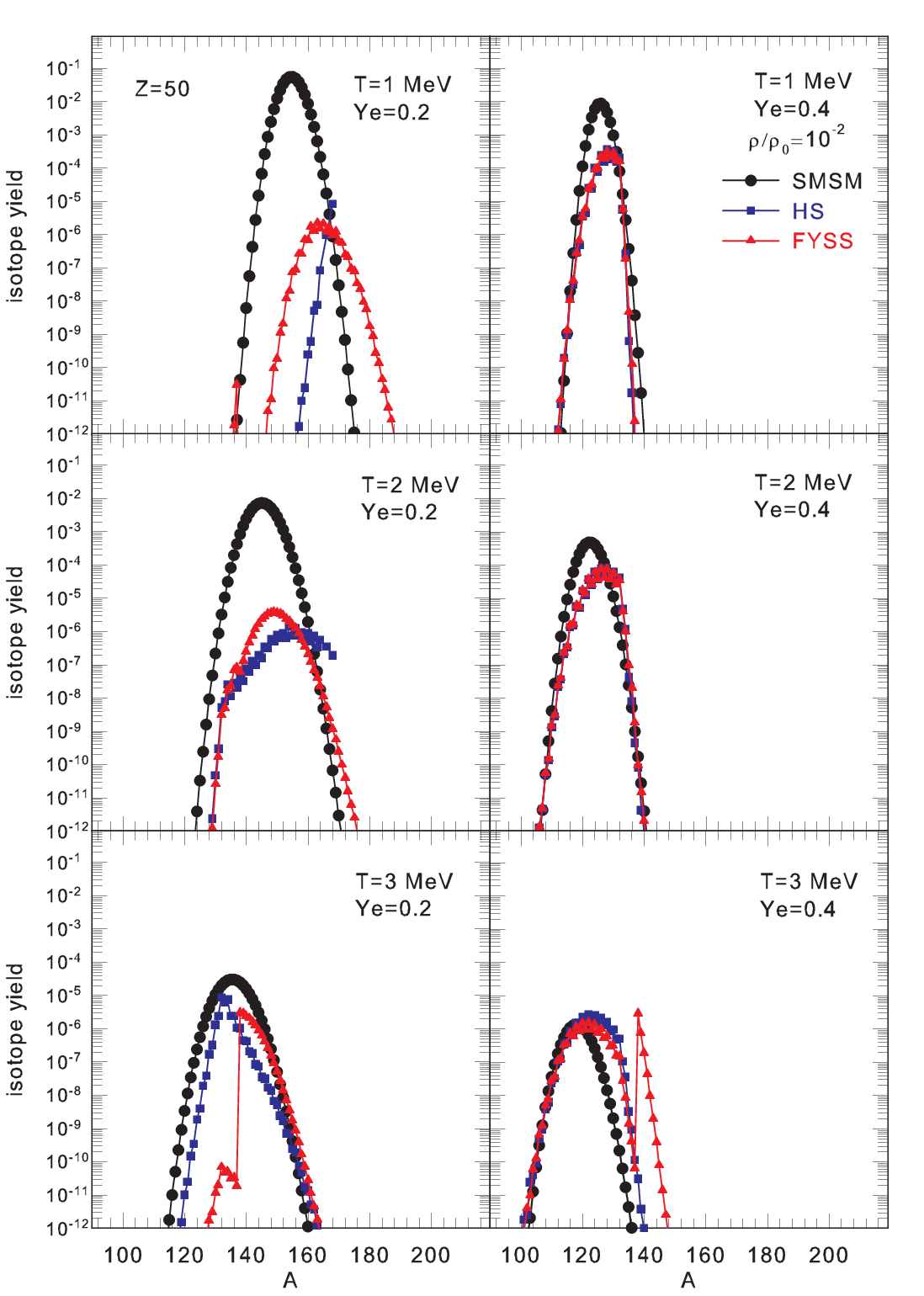}
\end{center}
\caption{\label{fig_yiso5}\small{Isotopic distributions of $Z=50$ fragments produced in matter with temperatures $T=1, 2$ and 3 MeV, electron
fractions $Y_e=0.2$ and 0.4, and  density $\rho/\rho_0=10^{-2}$. (Color version online.)
 }}
\end{figure}

\begin{figure}  
\begin{center}
\includegraphics[width=12cm,height=16cm]{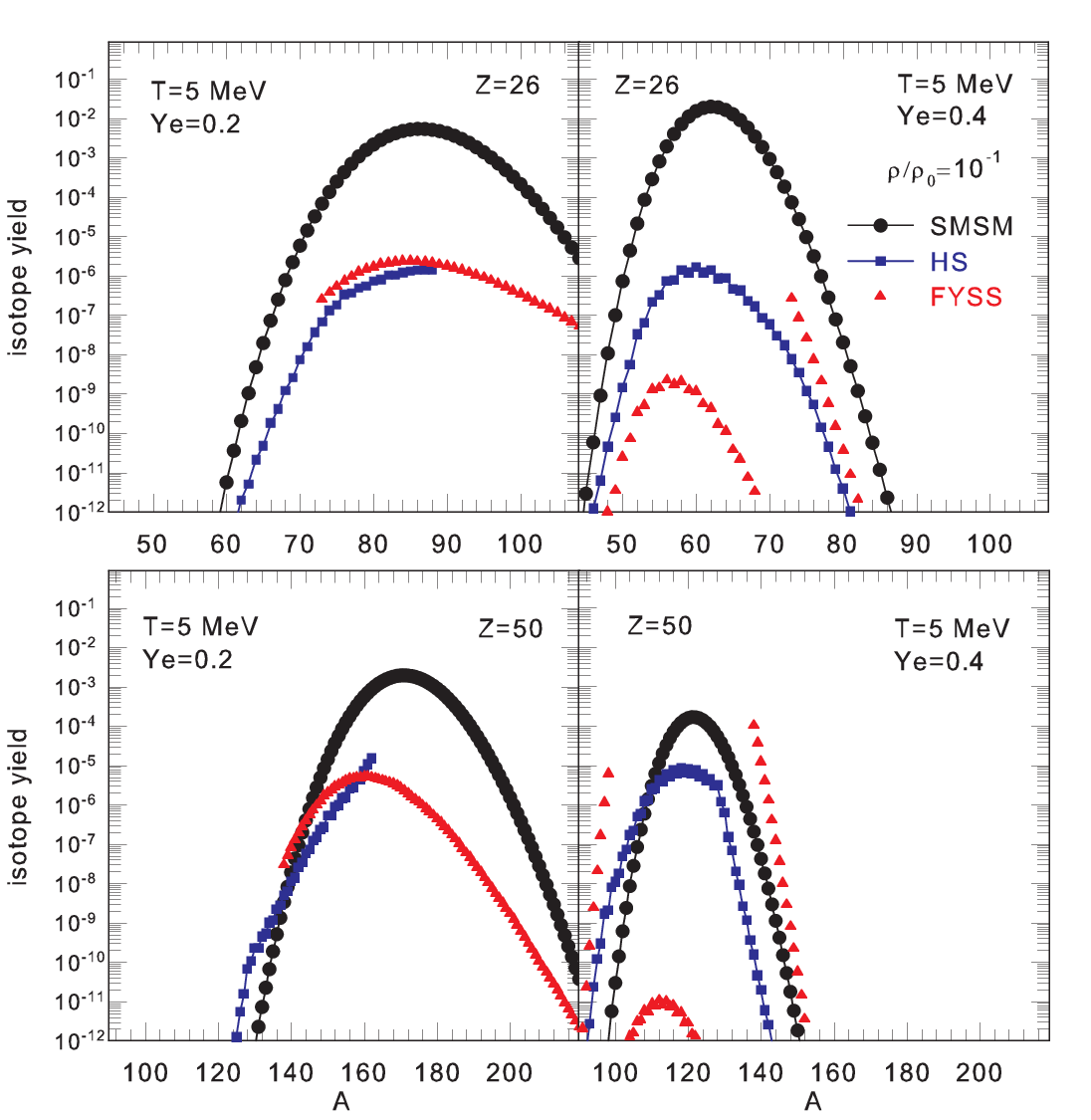}
\end{center}
\caption{\label{fig_yiso6}\small{Isotopic distributions of $Z=26$ and $Z=50$ fragments produced in matter with temperatures $T=5$ MeV, electron
fractions $Y_e=0.2$ and 0.4, and  density $\rho/\rho_0=10^{-1}$. (Color version online.)
 }}
\end{figure}

\begin{figure}  
\begin{center}
\includegraphics[width=12cm,height=16cm]{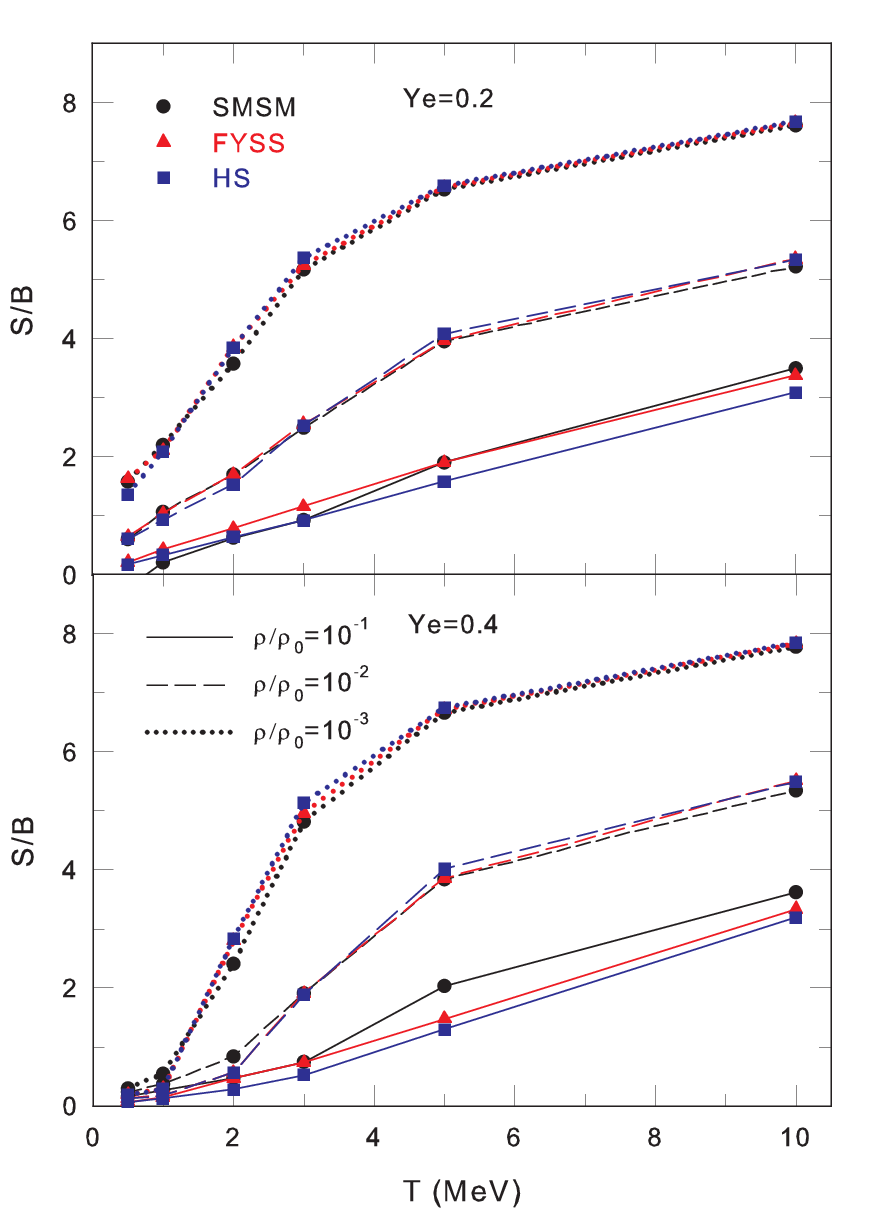}
\end{center}
\caption{\label{fig_entropy}\small{Comparison of SMSM, HS, and FYSS model results for the total nuclear entropy per baryon 
as a function of temperature. (Color version online.)
 }}
\end{figure}

Finally, we present thermodynamical characteristics of stellar matter 
predicted by the three models. In Fig.~\ref{fig_entropy}, 
we show the nuclear entropy per baryon as a function of temperature. As 
seen, the nuclear entropy per baryon is increasing with 
temperature, and all models give more or less similar results. This 
conclusion remains true if we extend  comparison to other similar models 
as in Refs. \cite{Japan,Janka2}. The largest differences occur at the 
highest density considered, $\rho/\rho_0 = 0.1$. SMSM 
tends to give the largest entropies for $T\geq 5$MeV because of entropy 
accumulation in internal excitation of fragments. In Fig.~\ref{fig_xheavy} 
one sees that the composition is dominated by heavy nuclei for these 
densities. Here we have to be more specific: Fig.~\ref{fig_aheavy} shows 
that $\langle A \rangle _{\rm heavy}$ drops below 100 for $T>3$ MeV and 
even below 20 for $T>5$ MeV. Thus the differences observed for the entropy 
for $T\geq 5$ MeV can be traced back to the different description of the 
free energies of low-mass nuclei in the three models. One also sees in 
Fig.~\ref{fig_xheavy} that in the SMSM the mass fraction of ``heavy'' 
nuclei remains close to one, even for temperatures of 10~MeV. This agrees 
with the finding that the entropy in SMSM is slightly increased, indicating 
a temperature dependence which favors intermediate mass nuclei in SMSM. In 
the other two models, which both use the internal partition function, 
a decrease of $X_{\rm heavy}$ is observed for temperatures larger than 
5~MeV and the results are more similar. 

\begin{figure} 
\begin{center}
\includegraphics[width=16cm,height=16cm]{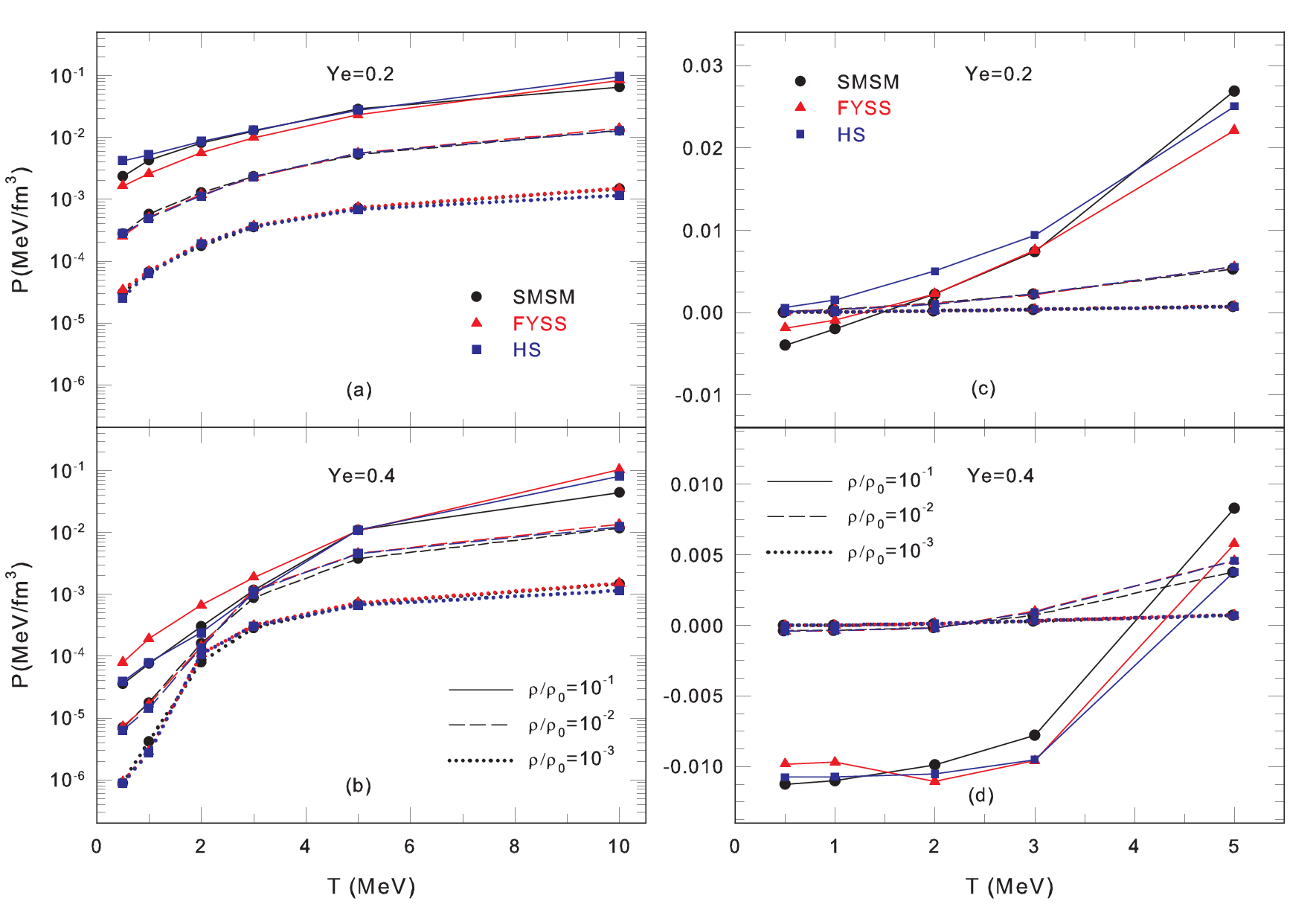}
\end{center}
\caption{\label{fig_pressure}\small{Comparison of SMSM, HS, and FYSS model results for the pure ((a)-(b)) 
and total ((c)-(d))
nuclear pressure as a function of temperature. (Color version online.)
 }}
\end{figure}

\begin{figure}   
\begin{center}
\includegraphics[width=12cm,height=16cm]{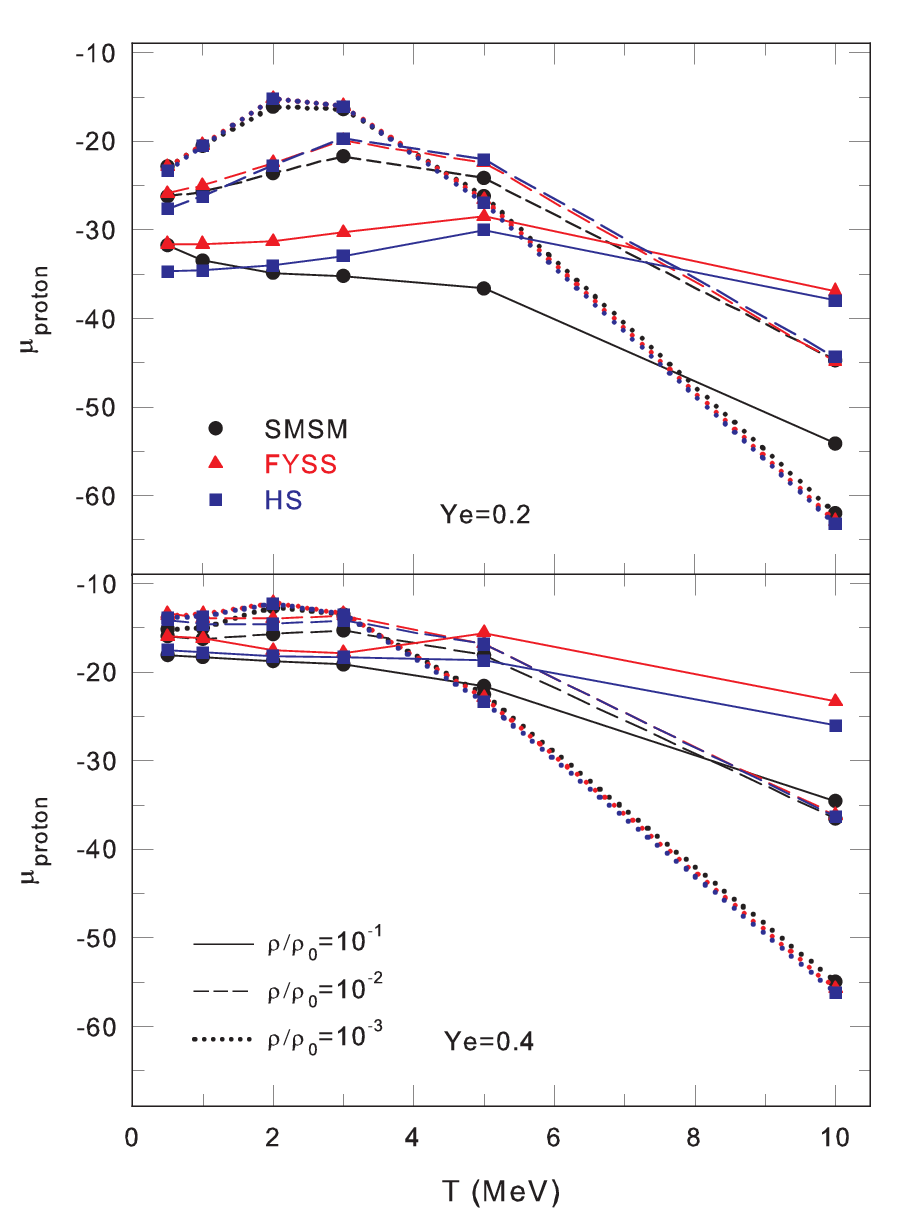}
\end{center}
\caption{\label{fig_mup}\small{The chemical potential of protons as a function of 
temperature. (Color version online.)
 }}
\end{figure}

\begin{figure}  
\begin{center}
\includegraphics[width=12cm,height=16cm]{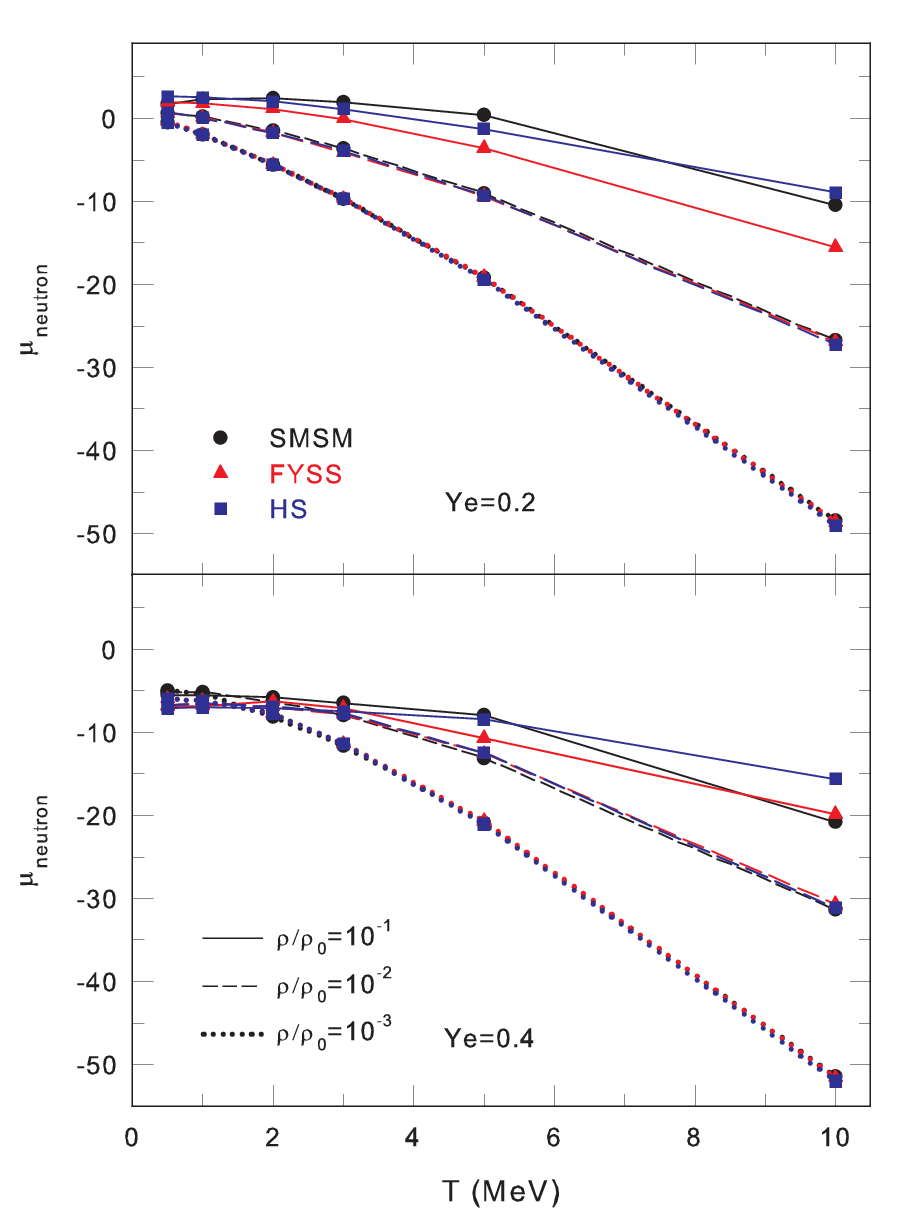}
\end{center}
\caption{\label{fig_mun}\small{The chemical potential of neutrons as a function of 
temperature. (Color version online.)
 }}
\end{figure}

In Fig.~\ref{fig_pressure}(a)-(b), we present the contributions to pressure 
caused by 
single nucleons and nuclei produced after clusterization of nucleons. 
In the following we call it pure nuclear pressure. The differences between 
the models are seen especially at the largest density. They are mainly 
caused by differences in fractions of nucleons and nuclei
discussed previously. 
The mean-field effects taken into account in FYSS and HS also influence
the results. In
addition, in the supernova environment free electrons existing  
together with nuclear clusters modify the Coulomb energy and lead to a 
negative Coulomb pressure \cite{BPS}. The physical reason is that  
electrons compensate the positive charge of nuclei, and the Coulomb energy 
of Wigner-Seitz cells decrease when the density increases. 
This Coulomb pressure can be calculated as: 
\begin{eqnarray}
 P_{C}=\rho_B \sum_{AZ} \rho_{AZ}  \frac{\partial F_{AZ}^C}{\partial \rho_B} ,
\end{eqnarray}
where the expression~(\ref{fazc}) for $F_{AZ}^C$ is used. 
The total nuclear pressure, which is the sum of the pure nuclear pressure and 
$P_{C}$, is demonstrated in Fig.~\ref{fig_pressure}(c)-(d) as a function of 
temperature. It is 
important that at $T \goo 5$MeV, when matter nearly completely dissociates 
into nucleons and lightest clusters, $P_C$ is close to zero. In this region 
the total nuclear pressure coincides with the pure nuclear pressure. The 
Coulomb pressure becomes very important when heavy clusters dominate in 
the system. One can see in Fig.~\ref{fig_pressure}(c)-(d) that at low 
temperatures and high density 
the total nuclear pressure may be negative. In this case the nuclear 
clusterization is favorable for the collapse, and
all three models show a similar behavior\footnote{Note that in Fig.~4 of Ref.~\cite{Botvina10} 
an inconsistent comparison was made between the SMSM pure nuclear pressure 
and the total nuclear pressure of the Shen model \cite{Shen}.}. 
However, the positive pressure 
of the relativistic degenerate electron Fermi-gas is
considerably (more than an 
order of 
magnitude) larger. Therefore the total pressure $P^{\rm tot}$
given by the sum of the total nuclear, electron and photon pressures in
such environment will always be positive and the condition of
thermodynamical stability 
( $\partial P^{\rm tot}/\partial \rho\geq 0$) will be fulfilled. 

The chemical potentials of both protons and neutrons are very important 
for all electron- and neutrino-induced reactions, which play an important 
role in supernova dynamics. They are shown in 
Figs.~\ref{fig_mup} and \ref{fig_mun}. One can see that 
results of SMSM, HS and FYSS EOS agree well and show the same behavior 
with density and temperature. However, there are deviations at the 
highest density $\rho/\rho_0=10^{-1}$ and high temperatures.  
They may be transformed into essential differences in mass and isotope 
distributions of produced fragments. It is interesting that $\mu_p$ obeys a 
rather similar trend for HS and FYSS, but in SMSM there is a strong decrease 
observed with increasing temperature. This should also be a result of the 
different description of heavy fragments, as discussed before. 
As one can see in Fig.~\ref{fig_xp}, HS and FYSS predict roughly 10\% of 
unbound protons for $T=10$~MeV, which is about one magnitude larger than in 
SMSM. This means that more protons are bound in SMSM nuclei. Also we can see 
$\mu_n$ of FYSS are  lower than in SMSM and HS at  
$\rho/\rho_0=10^{-1}$, high temperatures and $Y_e = 0.2$. This is because 
more neutrons are bound in FYSS nuclei, as shown in  Fig~\ref{fig_xn}. These 
differences may lead to significant effects in the weak reactions. 

\section{Conclusions}
\vspace{0.3cm}

We believe that the present comparative study of the three models, SMSM, HS and 
FYSS for the EOS will be very instructive for understanding
the 
differences and similarities in their predictions. These 
results can be used in hydrodynamical simulations of 
the collapse of massive stars and their subsequent explosions. 
The intervals of temperatures $T$=0.5--10 MeV and densities 
$\rho/\rho_0=10^{-1}-10^{-3}$ at electron fractions $Y_e$=0.2, 0.4 were 
under investigation. These conditions can occur during the main stages of 
the supernova process. All the three models give the detailed information 
about the nuclear mass and isotope distributions. In supernovae, the 
presence of heavy nuclei and their distributions 
are most important for electron captures and neutrino trapping during 
the collapse phase. We found that the width of the mass distributions can 
become even larger than 100 units, indicating substantial differences to the 
commonly used single nucleus approximation. It would be interesting to 
investigate the impact of the different EOSs and nuclear distributions 
on numerical simulations of the core-collapse supernovae in the future. 

On the other hand the investigated conditions are interesting for 
understanding the nuclear liquid-gas phase transition which can 
be investigated in laboratories, e.g., in multifragmentation reactions 
induced by heavy-ion collisions. 
Generally, we have concluded that at low density and high temperatures 
($T>2$ MeV), the three models give similar results for basic 
thermodynamical quantities like pressure, entropy, mass fractions of 
neutrons, protons, alphas and heavy nuclei, and chemical potentials of 
protons and 
neutrons. It is interesting that the mass distributions and other characteristics of ensemble nuclei differ from each other especially at high subnuclear densities. The differences result mainly from the dependence of bulk and surface energies on temperature and neutron-richness, and presence or absence of shell effects. We point especially at differences in isotope distributions of produced nuclei. Many isotopes are important for calculating the rate of weak reactions.
Since there is not enough data on the production of heavy nuclei in this 
region of the phase diagram, this striking difference between three models 
calls for new experimental investigations. Reactions of multifragmentation of excited neutron-rich nuclear systems into many fragments would be very suitable, since they provide a natural method for simulating conditions with low electron fraction. In supernova matter at subnuclear density we may expect nuclei with extremely high mass numbers and large isospin asymmetries. Therefore, these conditions should be initiated by collisions of large exotic and neutron-rich nuclei.

Regarding thermodynamic quantities like pressure or entropy, the largest 
differences occur at high densities $\rho/\rho_0 \sim 0.1$ and 
high temperatures $T > 5$~MeV. For such conditions intermediate mass and light 
nuclei are most important. The medium modification of such nuclei turns out 
to be the most important element of uncertainty in calculations of the 
supernova EOS, in addition to the bare nucleon-nucleon interactions. 
Part of the medium modification is 
due to Pauli-blocking, which can be calculated with quantum statistical 
approaches \cite{roepke11}. One could try to incorporate these binding 
energy-shifts phenomenologically in statistical models as mass corrections. 
One can use also knowledge obtained from experimental studies of nuclear 
multifragmentation 
reactions. For example, recent analyses of multifragmentation data 
give evidences for reduction of the symmetry energy coefficient in mass 
formula of the nuclei in such hot environment 
\cite{LeFevre,Souliotis,Ogul,Iglio,Hudan,Ogul09,Henzlova}. 
This may serve as a guideline for future microscopic theories. 
A theoretical approach in this direction is the so-called 
``generalized RMF'' model of Ref.~\cite{typel10}, which includes mean-field 
interactions of light nuclei. 

At low temperatures ($T<2$~MeV) but high densities another important 
uncertainty 
comes from medium modifications of the nuclear binding energies of heavy 
nuclei. 
Part of it refers to the evolution of the nuclear shell effects. It is 
for many years under theoretical \cite{Ignatiuk} and experimental 
\cite{K-H-Schmidt} investigations in nuclear reactions. 
This problem is also related to the general question on properties of nuclei 
for from stability, including the effect of temperature and surrounding 
medium of unbound nucleons and light clusters. 
Theoretical studies in this direction have been done recently 
within Skyrme-Hartree-Fock calculations \cite{newton09}, 
RMF models in the Hartree approximation \cite{gshen2010}, and within 
a simple model with two-body momentum-dependent interaction \cite{de2012}.  
However, these calculations are performed within the 
single nucleus approximation. Until now, no realistic
quantum many-body model has been calculated for all supernova conditions which 
takes into account distributions of different nuclei (light and heavy). At densities even higher than the densities we considered in our work, i.e., above $0.1 \rho_0$, additional effects like the formation of large-scale non-spherical structures (usually denoted as the pasta phases) may set in \cite{Lamb,newton09,watanabe11}. This is obviously beyond the scope of the present article, but part of these 
effects are already included in a simplified manner in some EOS models, 
e.g., in the FYSS model.

The comparative analysis of several models made in this paper show the necessity of theoretical and experimental studies of nuclei in the wide region of nuclear chart and their consistent treatment. For the basic input it is necessary to have experimental data on nuclei toward the neutron-rich side as much as possible. It is also important to study the nuclear compositions and modification of properties of nuclei at subsaturation densities in laboratory experiments, e.g., on heavy-ion collisions leading to multifragmentation reactions. We want to stress that excited nuclear matter created in these reactions in many respects is similar to supernova matter. As for theoretical side, it is necessary to construct a unified statistical approach taking into account ingredients which pass verification with experimental data at the corresponding thermodynamical conditions of nuclear matter. The binding energies and level densities of heavy nuclei at high temperature and density should utilize the medium modifications as predicted by microscopic theories. In addition, a consistent treatment of transformation of nuclei into 'pasta' and uniform matter is crucial in application of EoS for supernova simulations. Such efforts are indispensable for understanding both for astrophysics and nuclear physics.
\vspace{0.3cm}

{\bf  Acknowledgement}\\
\vspace{0.3cm}

N.B. and A.S.B. are supported by HIC for FAIR (LOEWE program) and grateful 
to 
Frankfurt Institute for Advanced Studies (FIAS) for support and hospitality. 
I.N.M. acknowledges partial support provided by grant NSH-215.2012.2 (Russia). 
M.~H. is supported by the High Performance and High Productivity Computing 
Project (HP2C), and the Swiss National Science Foundation (SNF) under 
project number No.\ 200020-132816/1. M.H.\ thanks Matthias Liebend\"orfer 
for useful discussions of the manuscript. J.S.-B. is supported by the German 
Research Foundation (DFG) within the framework of the excellence initiative 
through the Heidelberg Graduate School of Fundamental Physics. The authors 
are additionally supported by CompStar, a research networking program of the 
European Science Foundation (ESF). F.-K.T. and M.H. are
also grateful for 
participating in the EuroGENESIS collaborative research program of the ESF 
and the ENSAR/THEXO project. 
K.S., H.S. and S.Y. are partially supported by the Grant-in-Aid for Scientific Research 
on Innovative Areas (Nos. 20105004, 20105005), the Grant-in-Aid for the 
Scientific Research (Nos. 19104006, 21540281, 22540296, 24244036) and 
the HPCI Strategic Program from the Ministry of Education, Culture, Sports, 
Science and Technology (MEXT) in Japan. 
S.F. is grateful to FIAS for generous support. 
S.F. is supported by the Japan Society for the Promotion of Science 
Research Fellowship for Young Scientists (24E79). A part of the numerical calculations were carried out on SR16000 at  YITP in Kyoto University.
K.S. acknowledges the usage of the supercomputers at Research Center 
for Nuclear Physics (RCNP) in Osaka University, The University of Tokyo, 
Yukawa Institute for Theoretical Physics (YITP) in Kyoto University and 
High Energy Accelerator Research Organization (KEK). 
K.S. is grateful to the organizers of NUFRA2011, where this project was 
initiated, for fruitfully discussions and extensive collaborations afterwards.

\newpage

\end{document}